%% file: main.tex
\documentclass[sigconf]{acmart}
\AtBeginDocument{%
  }

\copyrightyear{2026}
\acmYear{2026}
\setcopyright{cc}
\setcctype{by}
\acmConference[ICSE '26]{2026 IEEE/ACM 48th International Conference on Software Engineering}{April 12--18, 2026}{Rio de Janeiro, Brazil}
\acmBooktitle{2026 IEEE/ACM 48th International Conference on Software Engineering (ICSE '26), April 12--18, 2026, Rio de Janeiro, Brazil}
\acmPrice{}
\acmDOI{10.1145/3744916.3773177}
\acmISBN{979-8-4007-2025-3/26/04}

\input{macros}

\begin{document}

\title{Energy-Efficient Software Development: A Multi-dimensional Empirical Analysis of Stack Overflow}

\author{Bihui Jin}
\affiliation{%
  \institution{University of Waterloo}
  \city{Waterloo}
  \country{Canada}}
\email{bihui.jin@uwaterloo.ca}

\author{Heng Li}
\affiliation{%
  \institution{Polytechnique Montréal}
  \city{Montréal}
  \country{Canada}}
\email{heng.li@polymtl.ca}

\author{Pengyu Nie}
\affiliation{%
  \institution{University of Waterloo}
  \city{Waterloo}
  \country{Canada}}
\email{pynie@uwaterloo.ca}

\author{Ying Zou}
\affiliation{%
  \institution{Queen's University}
  \city{Kingston}
  \country{Canada}}
\email{ying.zou@queensu.ca}

\renewcommand{\shortauthors}{Jin et al.}

\begin{abstract}
Energy consumption of software applications has emerged as a critical concern for developers to contemplate in their daily development processes.
Previous studies have surveyed a limited number of developers to understand their viewpoints on energy consumption.
We complement these studies by analyzing a meticulously curated dataset of 1,193 Stack Overflow (SO) questions concerning energy consumption.
These questions capture real-world energy-related challenges in practice.
To understand practitioners' perceptions, we investigate the intentions behind these questions, semantic topics, and associated technologies (e.g., programming languages).
Our results reveal that: (i)~the most prevalent energy consumption topic is about balancing \gps usage;
(ii)~efficiently handling data is particularly challenging, with these questions having the longest response times;
(iii)~practitioners primarily ask questions to understand a concept or API related to energy consumption;
and (iv)~practitioners are concerned about energy consumption across multiple levels---hardware, operating systems, and programming languages---during energy efficient software development.
Our findings raise awareness about energy consumption's impact on software development. \Revision{We also derive actionable implications for energy optimization at different levels (e.g., optimizing API usage or hardware accesses) during energy-aware software development}. 
\end{abstract}

\maketitle

\input{Text/Intro}
\input{Text/Related_Work}

\input{Text/Exp}
\input{Text/Results}
\input{Text/Impli}
\input{Text/Threats}
\input{Text/Concl}
\begin{acks}
We thank Dr. Stefanos Georgiou for his helpful discussions and feedback on the early stages of this work, and the anonymous reviewers for their insightful comments and feedback. 
\end{acks}

\clearpage
\balance
\bibliographystyle{ACM-Reference-Format}
\bibliography{bib}

\end{document}

%% file: macros.tex
\usepackage{xspace}
\usepackage[table,xcdraw]{xcolor}
\usepackage{graphicx}
\usepackage[normalem]{ulem} 
\usepackage{amsmath}
\usepackage{amsfonts}
\usepackage{enumitem}
\usepackage{textcomp}
\usepackage[para,online,flushleft]{threeparttable}
\usepackage{hyperref}
\hypersetup{linkcolor=black,citecolor=black,anchorcolor=black,filecolor=black,menucolor=black,runcolor=black,urlcolor=black,hidelinks}
\usepackage{breakurl}

\usepackage{booktabs}
\usepackage{multirow}
\usepackage{makecell}
\usepackage{ragged2e}

\usepackage{caption}
\usepackage{subcaption}

\usepackage{listings}

\usepackage{tikz}
\tikzset{%
    baseline,
    inner sep=2pt,
    minimum height=12pt,
    rounded corners=2pt  
}
\usetikzlibrary{calc}
\usetikzlibrary{shapes.geometric}
\usetikzlibrary{decorations.pathreplacing}
\usetikzlibrary{positioning}
\usetikzlibrary{shapes}

\usepackage{datetime} 
\usepackage{tcolorbox} 
\usepackage{geometry}
\usepackage{fancyhdr}
\usepackage{color}
\usepackage{balance}
\usepackage{everypage}
\usepackage{todonotes}
\usepackage{subcaption}
    
\useunder{\uline}{\ul}{}

\newcommand{\bihui}[1]{\textcolor{red}{[BJ: #1]}}

\newcommand{\Revision}[1]{\textcolor{black}{#1}}

\newcommand{\code}[1]{\mbox{
    \ttfamily
    \tikz \node[anchor=base,fill=black!12]{#1};
}}
\newcommand{\XSpace}[1]{}
\newcommand{\XComment}[1]{}

\makeatletter
\newcommand{\DefMacro}{\@ifstar\@DefMacroAllowRedefine\@DefMacro}
\newcommand{\@DefMacro}[2]{\expandafter\newcommand\csname rmk-#1\endcsname{#2}}
\newcommand{\@DefMacroAllowRedefine}[2]{\expandafter\providecommand\csname rmk-#1\endcsname{} \expandafter\renewcommand\csname rmk-#1\endcsname{#2}}
\makeatother
\newcommand{\UseMacro}[1]{\csname rmk-#1\endcsname}

\newcommand{\MyPara}[1]{\noindent\textbf{#1}.}

\newcommand{\InputWithSpace}[1]{\bgroup\def\arraystretch{1.1}\input{#1}\egroup}
\newcommand{\Code}[1]{{\ifmmode{\mathtt{#1}}\else$\mathtt{#1}$\fi}}
\newcommand{\CodeIn}[1]{{\ifmmode{\mathtt{#1}}\else$\mathtt{#1}$\fi}}

\newcommand{\concept}{{\tt CONCEPTUAL}\xspace}
\newcommand{\api}{{\tt API USAGE}\xspace}
\newcommand{\disc}{{\tt DISCREPANCY}\xspace}
\newcommand{\review}{{\tt REVIEW}\xspace}

\newcommand{\gps}{\textit{Positioning}\xspace}
\newcommand{\resource}{\textit{Computing Resource}\xspace}
\newcommand{\mobile}{\textit{Mobile Device}\xspace}
\newcommand{\sensor}{\textit{Sensor Timing}\xspace}
\newcommand{\data}{\textit{Datum Handling}\xspace}
\newcommand{\trans}{\textit{Data Transmission}\xspace}
\newcommand{\polling}{\textit{Polling}\xspace}
\newcommand{\thread}{\textit{Thread}\xspace}

\newcommand{\os}{\textit{OS}\xspace}
\newcommand{\pl}{\textit{PL}\xspace}
\newcommand{\hw}{\textit{HW}\xspace}

\newcommand*{\conclusionbox}[1]{%
    \vspace{2mm}
    \noindent
    \fcolorbox{black}{gray!10}{%
        \parbox[b]{0.96\columnwidth}{%
            \emph{#1}
        }
    }
    \vspace{-2mm}
}

\DeclareMathAlphabet{\mathbbmsl}{U}{bbm}{m}{sl}
\newcounter{observations}
\newcommand{\sopost}[1]{\href{https://stackoverflow.com/questions/#1}{post #1}}

%% file: Text/Intro.tex
\section{Introduction}\label{sec:Intro}
Technological advancement has led to increased energy consumption~\citep{Jin}.
Particularly, in the era of artificial intelligence, the energy consumption of Information and Communication Technology (ICT) is projected to surge by 822.79\% from 2001 to 2030, reaching 17959.11 TWh per year~\citep{Trends2030}.
In 2020, ICT accounted for up to 7\% of global electricity use~\citep{2030} and 2.1\% to 3.9\% of global greenhouse gas emissions~\citep{emissions}.
By 2030, data centers alone are projected to consume 10\% of the world's electricity~\citep{data2030}.

Software plays a crucial role in ICT energy consumption~\citep{software2}.
Energy-related issues affect every phase of the software lifecycle, including design, implementation, testing, and maintenance~\citep{software1}.
Energy-efficient software can save energy, extend battery life, and enhance user experience~\citep{Pang}, helping mitigate the growing trend of global electricity consumption.
Practitioners now view energy efficiency as a key software property and have demonstrated willingness to learn about energy issues in software development~\citep{Manotas}.

A few empirical studies have been conducted to understand practitioners' perceptions of energy consumption.
For instance, prior studies~\citep{Pang, Manotas} conduct online surveys with 122 and 464 practitioners, respectively, who self-identify as experienced developers and testers across various application domains.
The studies find that 65\% of practitioners on Reddit recognize energy usage as a crucial factor in software quality~\citep{Pang}, and many are willing to sacrifice other requirements to reduce energy consumption~\citep{Manotas}.
However, these studies only survey opinions rather than the concrete barriers practitioners face when developing energy-efficient software.
The most similar work to ours is by~\citet{Pinto}, who study 325 Stack Overflow (SO) questions from 2008 to 2013 and identify five primary themes of energy-related questions (e.g., measurements and code design).
Given rapid technological changes (e.g., IoT devices), practitioners likely face different challenges today than a decade ago, making the earlier studies less relevant to current practitioners.

\Revision{We follow the same methodology outlined in~\citet{Pinto} to collect energy-related posts. Through keyword search followed by manual verifications, we curated a dataset of 1,193 energy-related SO questions, which contains \textbf{larger and more recent} questions compared to the work by Pinto et al. (325 questions by 2013). 
Inspired by recent work on analyzing SO posts~\cite{Barua, Yang, RS, blockchain}, we perform \textbf{LDA topic modeling instead of thematic analysis} to obtain low-level and actionable semantic topics (e.g., \emph{positioning} or \emph{data transmission}) in these questions. 
In addition, we investigate \textbf{new dimensions} of these energy-related questions, including their intentions (e.g., to understand a concept) and the associated technologies (e.g., an operating system). Furthermore, we analyze the \textbf{evolution} of these energy-related questions over the years. 
}
We formulate three research questions to guide our study:

\begin{itemize}[leftmargin=*, itemsep=1ex]
    \item \textbf{RQ1: What are the topics of the energy-related questions and their difficulty?} 
    
    The energy-related questions asked by practitioners on SO are diverse in nature.
    To understand the characteristics of SO questions, we apply topic modeling to analyze them and group them into topics.
    We identify eight recurring topics related to energy issues: \gps, \resource, \mobile, \sensor, \polling, \data, \trans, and \thread.
    Questions about \texttt{GPS} (e.g., location tracking) are the most common, while questions about \texttt{Polling} are the most challenging, requiring the longest time to receive community responses.
\end{itemize}  

\begin{itemize}[leftmargin=*, itemsep=1ex]
    \item \textbf{RQ2: What are the intentions behind energy-efficient development  questions?}
    
    Practitioners have diverse intentions when posting energy-related questions, such as finding the root causes of rising energy consumption.
    Understanding the intentions provides insights into the challenges associated with energy consumption in software development~\citep{Allamanis}.
    We observe that nearly half of the energy-related questions are mainly \Revision{concept}-oriented, covering background knowledge of APIs and programming concepts like design patterns.
    These questions are frequently accompanied by \api questions about learning how to use an API.
    The third most common intention is \disc, where practitioners struggle to resolve discrepancies between the profiling results and their expectations regarding energy consumption.
\end{itemize}  

\begin{itemize}[leftmargin=*, itemsep=1ex]
    \item \textbf{RQ3: What technologies are concerned in energy-efficient development?}
    
    Energy-related questions span several levels of technologies, including operating systems (OS), programming languages (PL), and hardware (HW),
    usually identified by tags (e.g., \texttt{android}, \texttt{java}, and \texttt{raspberry-pi}).
    We analyze the tags attached to each SO post to categorize energy-related concerns.
    Through our analysis of tag categories, we provide suggestions for allocating technical support to help practitioners solve energy-related problems.
\end{itemize}

\noindent \Revision{This work aims to understand practitioners' challenges in energy-aware software development through examining SO questions where practitioners express their challenges. 
Each SO question carries information from several mutually complementary perspectives about practitioners' challenges: the semantical meaning (topic) of the question (RQ1), the intentions of the question (RQ2), and the associated technology (RQ3), motivating our multi-dimensional investigation.} 
The main contributions of this study include:
 
\begin{itemize}[topsep=3pt,itemsep=1ex,partopsep=0ex,parsep=0ex,leftmargin=*]

    \item We provide a carefully curated dataset of 1,193 SO posts (2008 - 2024) on energy-related concerns, 
    analyzed via a combination of manual, topic modeling, and LLM-based efforts. Our public dataset will be a valuable resource for future research on energy-efficient software development.

    \item Our approach and empirical findings gain insights on energy-related SO posts in three dimensions: topics, intentions, and associated technologies, extending Pinto et al.'s analysis that focuses only on the topics of 325 posts (from 2008–2013). We also discuss the potential root causes leading to developers' questions. 

    \item \Revision{Our analyses of topics, intentions and associated technological shifts provide insights on the common types of questions developers ask (e.g., questions about API usages) and the direction for support in application level and specialized device contexts.}


    \item 
    \Revision{We derive actionable implications at  API-level and low-level OS- and hardware-level energy optimizations to improve effectiveness in energy-efficient software development.}
\end{itemize}

Our dataset and experiment scripts are open-source on GitHub~\cite{RepPackage}.


%% file: Text/Related_Work.tex
\section{Related Work}\label{sec:RW}


\subsection{Energy-Efficient Software Development}
\citet{Cruz19Catelog} create a catalog of 22 design patterns for improving mobile app energy efficiency based on analyzing 1,027 Android and 756 iOS apps. They find Android developers are more aware of energy issues than iOS developers and provide actionable guidelines for improving energy consumption.
%
\citet{Bao16Android} study 468 power management commits from 154 Android apps, categorizing them into six activities, including power adaptation, consumption improvement, and wake lock optimization. Their findings highlight how different app categories prioritize different power management strategies.
%
\citet{Moura15MiningEnergy} identify 12 themes in 371 energy-aware commits across 317 applications. They find that developers struggle to predict the energy impact of code changes and that energy-saving techniques may compromise other quality attributes. 
Our work complements code- and commit-based studies by examining developers' questions, revealing practical knowledge gaps that exist in practice.

\citet{Pang} survey over 100 practitioners and find that 86\% lacked awareness of energy efficiency best practices and could not identify the causes of high energy consumption. Their study reveals that practitioners struggle to implement energy-efficient solutions due to insufficient information, tools, and infrastructure.
%
\citet{Manotas} analyze survey responses from 464 experienced practitioners at major tech companies and find that developers are willing to sacrifice other requirements to reduce energy usage but struggle to diagnose energy issues. 
While these survey-based studies provide valuable insights, our work complements them by analyzing real-world questions from SO, revealing the specific challenges practitioners face during actual development.

\citet{Pinto} analyze 325 SO questions from 2008 to 2013 related to energy consumption, finding that such questions increased yearly and that mobile development accounted for 25\% of the tags.
Our work extends this research with a larger dataset (1,193 questions from 2008 to 2024), providing an updated view of practitioners' challenges. Unlike \citet{Pinto}, who focus primarily on question topics, we analyze posts not only by topics but also along two additional dimensions (intent-based and technology-based), offering a more comprehensive insight into energy-efficient software development challenges.

\begin{figure*}[t]
\centering
    \includegraphics[width=\linewidth]{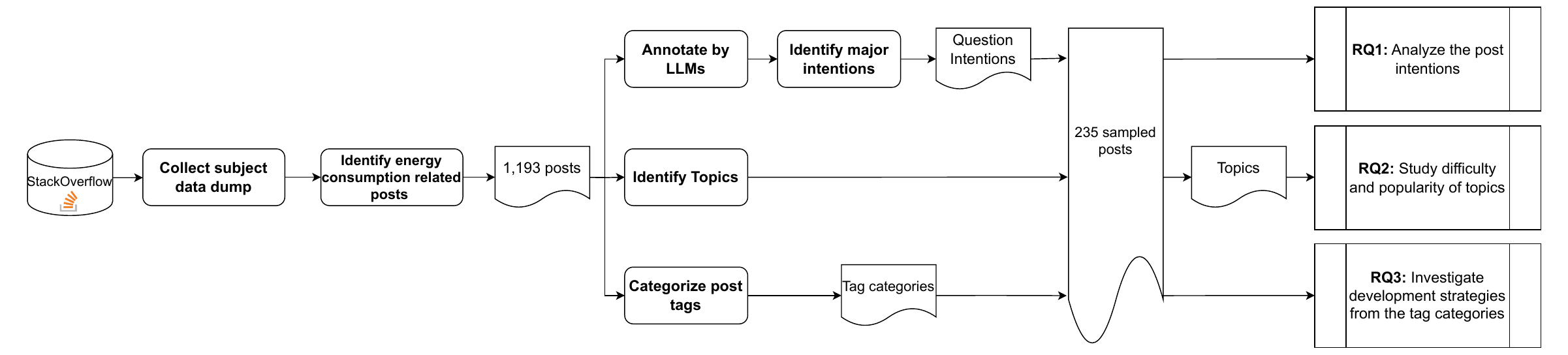}
    \vspace{-8mm}
    \caption{An overview of our experiment design for collecting, filtering, and processing SO posts.}
    \label{fig:flowchat}
    \vspace{-4mm}
\end{figure*}

\subsection{Analysis of Developer Forum Posts}

\citet{category} inspect and classify 1,000 Android-related SO posts into seven intention categories: {\tt API Usage}, {\tt Conceptual}, {\tt Discrepancy}, {\tt Errors}, {\tt Review}, {\tt API change}, and {\tt Learning}. They find {\tt API Usage} is the most common intention, followed by {\tt Discrepancy} and {\tt Conceptual}. We adopt this classification scheme in our study but find that for energy-related questions, {\tt Conceptual} questions are most common, highlighting the knowledge gap in energy-efficient development.


Previous studies have explored trends and topics in developer discussions using both temporal analysis and topic modeling.
\citet{Venkatesh} analyze 92,471 discussions about 32 Web APIs from developer forums, identifying five dominant patterns in how topics change over time. Their temporal analysis approach inspires our examination of how energy-related concerns have evolved over the years.
In addition, several studies have applied topic modeling to analyze SO posts: \citet{blockchain} cluster blockchain discussions; \citet{RS} summarize mobile-related questions; \citet{Yang} cluster security-related questions; and \citet{Barua} investigate primary discussion topics and trends over time. Following these approaches, we apply LDA~\cite{Blei}, a mature topic modeling technique, to identify eight recurring topics in energy-related questions, including \gps, \resource, and \data.

Our work uniquely combines these approaches to analyze energy-related SO posts. We adopt \citet{category}'s intention classification, use LDA for topic modeling like \citet{blockchain, RS, Barua}, and analyze energy-related technologies to build a multi-dimensional view of energy-related concerns
that are not captured in previous studies.

%% file: Text/Exp.tex
\section{Study Design} \label{sec:RM}

In this section, we present our data collection, pre-processing, and analysis steps. Figure~\ref{fig:flowchat} shows an overview of our approach.

\subsection{Subject Dataset}
\begin{figure}[ht!]
\centerline{\includegraphics[width=1.7\textwidth,height=0.6\textheight,keepaspectratio]{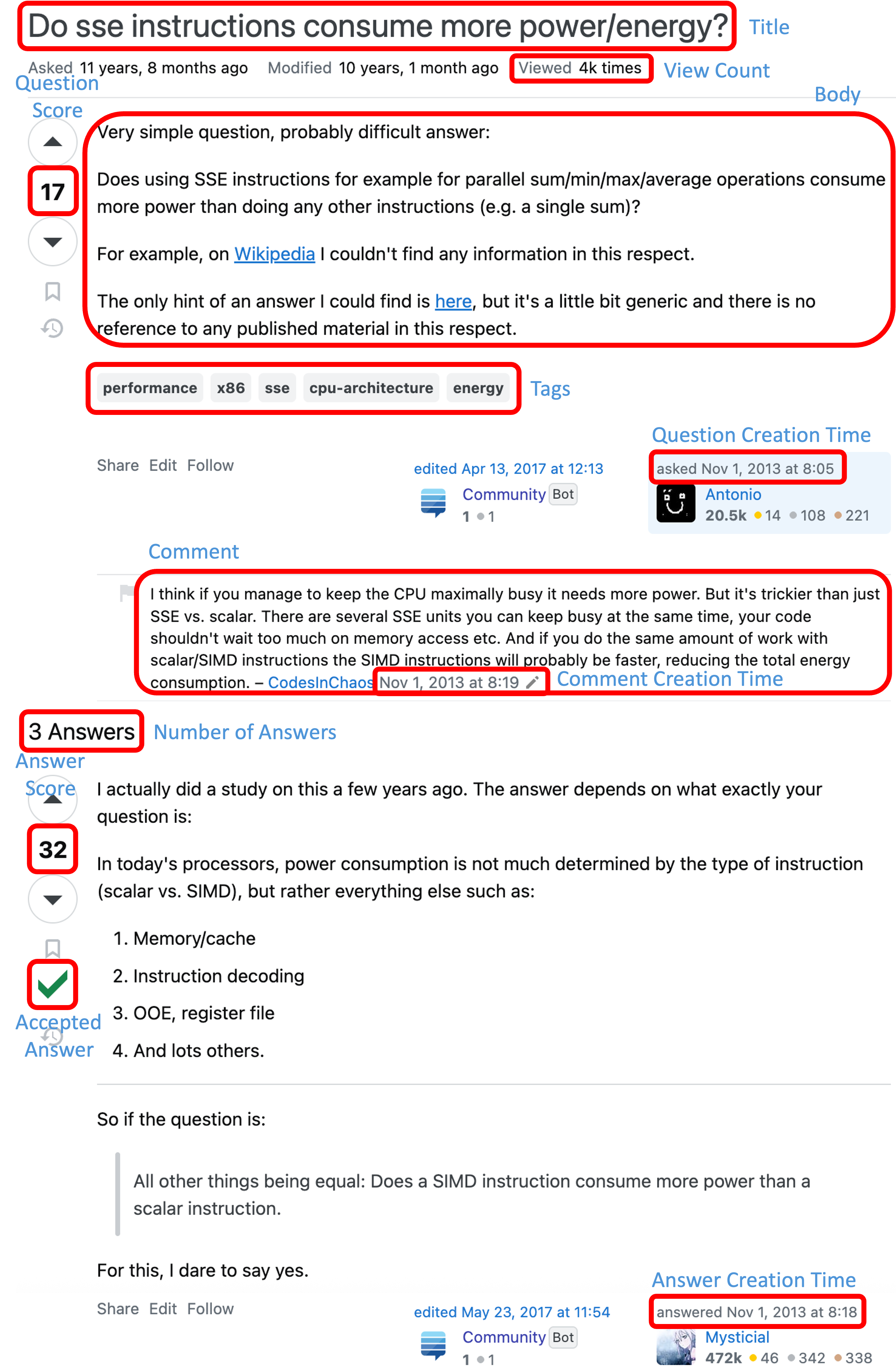}}
\vspace{-4mm}
\caption{An annotated screenshot of a Stack Overflow post.}
\label{fig:post}
\vspace{-4mm}
\end{figure}

We download the latest Stack Overflow official data dump\footnote{\url{https://archive.org/details/stackexchange_20241231}} at the time of conducting this work, which contains 23,709,404 posts from January 2008 to December 2024.
%
Figure~\ref{fig:post} shows a sample SO post~\cite{so19722950}. We extract the following information from each post:
\begin{itemize}[leftmargin=*]
    \item Title, which summarizes the question being asked or the topic being discussed;
    \item View count, showing how many times the post has been viewed;
    \item Body, containing details of the question or discussion;
    \item Tags, which are labels that have been assigned to the post according to the areas of the question;
    \item Question score, assigned by users based on its usefulness and clarity;
    \item Question creation time, when the question was first posted;
    \item Comments (to the question), including suggestions (that are not sufficient to be answers), clarifications, and relevant information added by other users;
    \item Comment creation time (of the first comment if it exists), which indicates how quickly it is to receive initial help to the question;
    \item Answers, which are the solutions to the question; an answer can be marked as ``accepted'' if it is verified to solve the question;
    \item Answer score (of the accepted answer if it exists), assigned by users based on its correctness and relevance;
    \item Answer creation time (of the accepted answer if it exists), which indicates how quickly the question is solved.
\end{itemize}

\subsection{Identifying Energy-Related SO Posts}\label{identifyPost}

We use a two-phase approach to identify SO posts related to energy consumption: an automated filtering phase to search keywords related to energy consumption in the question title and body, followed by a manual confirmation phase.

We first exclude code snippets from the posts surrounded by \textless code\textgreater\textless/code\textgreater  html tags, as they may contain incidental matches that do not reflect the intent of the questioner.
Then, we search posts based on energy-related keywords.
\Revision{The original eight keywords from Pinto et al. are proposed in 2014, whereas new energy-related keywords may surface over the years.
Thus, we manually inspect statistically significant samples and summarize four new keywords following Pinto et al.'s methodology, resulting in twelve keywords in total:}
\%energy consum\%, \%power consum\%, \%energy efficien\%,
\%power efficien\%, \%energy sav\%, \%power sav\%,
\%energy us\%, \%power us\%,
\%energy econ\%, \%power econ\%, \%sav\% energy\%,
\%sav\% power\%. 
The character `\%' enables fuzzy matching using the stems of the keywords. This search yields 2,629 potential posts.

We manually review all 2,629 posts to eliminate false positives---posts that mention energy-related keywords but do not actually seek to address energy consumption issues. As an example, the questioner may mention ``Energy Consumption'' in the title but not seek for energy-efficiency solutions~\cite{so13476223}.
%
\Revision{The labelling process involves two phases: 
\begin{enumerate}[topsep=3pt,itemsep=.5ex,partopsep=0ex,parsep=0ex,leftmargin=*]
\item Two annotators (PhDs in SE) independently label the first half of the questions, following a closed coding procedure~\citep{Seaman}. The Cohen's Kappa score~\cite{kappa} is 0.55 after this step, indicating a moderate level of agreement. Then, the two annotators discuss the labels of the first half and achieve consensus for each question, establishing a common understanding.
\item The two annotators then independently label the second half of the questions. The Cohen's Kappa score is 0.69 (substantial level) for the second half. Finally, the two annotators discuss the labels of the second half and achieve consensus.
\end{enumerate}
If any conflicts persist, a third annotator (who is a Software Engineering Professor and also an author of this paper) arbitrate remaining conflicts.
This process yields our final dataset of 1,193 posts with 1,478 answers.
}

\subsection{Pre-processing Posts} \label{preprocessing}
We pre-process the posts to prepare them for topic modeling, following these steps:

 \begin{itemize}[topsep=3pt,itemsep=.5ex,partopsep=0ex,parsep=0ex,leftmargin=*]
    \item \MyPara{Title and Body Composition}
    We concatenate each post's title and body to provide complete context for the model.
    
    \item \MyPara{Tokenization, Stemming, and Lemmatization}
    We process the text with {\tt spaCy} ({\tt en\_core\_web\_trf})~\cite{JUGRAN21Spacy,Hu2022DataDI} to tokenize the text, perform stemming (reducing words to their basic form), and lemmatization (normalizing word tense).
    
    \item \MyPara{Stop Words Removal}
    We remove common stop words, such as ``the'', ``an'', ``by'', ``is'', and ``what''. We also inspect the top 100 most frequent words in our dataset and filter out additional stop words that are irrelevant for our study, such as ``like'', ``thank''.
 
    \item \MyPara{Bag of Words}
    We create a Bag of Words representation using {\tt Gensim}'s {\tt doc2bow} function~\citep{rehurek}, which maps each document to a dictionary from token ids to counts, preserving word frequency while disregarding syntax and word order.
\end{itemize}

\input{Tables/lda_Topics}

%% file: Tables/lda_Topics.tex
\begin{table}
\caption{The LDA dominant topics in SO energy-related posts.}
\vspace{-3mm}
\label{tab:RQ2}
\centering
\resizebox{\linewidth}{!}{
\begin{tabular}{|c|l|}
\hline
\rowcolor[HTML]{C0C0C0} 
Topic                & \multicolumn{1}{c|}{\cellcolor[HTML]{C0C0C0}Keywords}                                                                            \\ \hline
Positioning                & \begin{tabular}[c]{@{}l@{}}energy, battery, location, usage (power, \\ app, consumption, android, use, application)\end{tabular} \\ \hline
\rowcolor[HTML]{C0C0C0} 
Computing Resource & \begin{tabular}[c]{@{}l@{}}cpu, gpu, core (power, consumption, \\ energy, time, usage, device, code)\end{tabular}                \\ \hline
Mobile Device      & \begin{tabular}[c]{@{}l@{}}power, consumption, android, device, \\ mode, application save, phone (code, app)\end{tabular}        \\ \hline
\rowcolor[HTML]{C0C0C0} 
Sensor Timing      & \begin{tabular}[c]{@{}l@{}}sensor, app, time, find, run\\ (power, consumption, use, user, code)\end{tabular}                     \\ \hline
Polling            & \begin{tabular}[c]{@{}l@{}}code, use, clock, sleep, module, wake \\ (power, consumption, mode, device)\end{tabular}              \\ \hline
\rowcolor[HTML]{C0C0C0} 
Datum Handling     & \begin{tabular}[c]{@{}l@{}}memory, kwh, try, datum, image\\ (energy, power, consumption, time)\end{tabular}                      \\ \hline
Data Transmission  & \begin{tabular}[c]{@{}l@{}}file, server, message (energy, device, app, \\ power, consumption, datum, try)\end{tabular}           \\ \hline
\rowcolor[HTML]{C0C0C0} 
Thread             & \begin{tabular}[c]{@{}l@{}}user, thread (power, code, use, mode,\\ application, try, consumption, datum)\end{tabular}            \\ \hline
\end{tabular}
}
\vspace{-6mm}
\end{table}

%% file: Text/Results.tex
\section{Results and Findings}\label{sec:ER}

In this section,
we describe the motivation, the approach, and our findings for each of our three research questions.

\input{Text/RQ1}
\input{Text/RQ2}
\input{Text/RQ3}

%% file: Text/RQ1.tex
\subsection{RQ1: What are the topics of the energy-related issues and their difficulty?}\label{rq1}

\subsubsection{Motivation}

The energy-related SO questions raised by practitioners cover diverse topics. 
Understanding the topics helps uncover their predominant concerns. 
We use topic modeling to cluster diverse energy-related concerns and analyze each topic's popularity and difficulty in receiving community support. 
This provides insights into the challenges developers face when working on energy-efficient software. 


\input{Tables/Topic_def}

\subsubsection{Approach} 

To discern question patterns, we perform the following steps:

\MyPara{Topic Modeling} \label{lda}
To study the most significant concerns regarding energy consumption, we utilize topic modeling to group the related posts into a topic. 
Specifically, we apply Latent Dirichlet Allocation (LDA)~\citep{LDA} from the {\tt Gensim} package~\citep{rehurek}. 
LDA is a generative statistical model that infers common topics based on the probability of the distribution of discrete words in a corpus, without predefined taxonomies~\citep{Blei, Griffiths}.
\Revision{LDA topic modeling is widely used in the analysis of SO posts~\cite{Barua,Yang, RS, blockchain}. 
Instead of manual coding, we use LDA to uncover latent topic structures and achieve unbiased and reliable clustering results.
}
%
LDA generates topics as collections of keywords with significance percentages. For example, in the set (0.023*cpu + 0.023*power + 0.016*consumption + 0.012*gpu...), ``cpu'' is the most significant word. 
Keywords can appear in multiple topics with different significance levels.

However, LDA requires a hyperparameter of the number of topics, the optimal number of which is unknown before running the analysis.
We train the model for 100 iterations with 1--100 topics and determine the optimal number of topics using \emph{Jaccard similarity} and \emph{topic coherence} metrics.
For other parameters, such as $\alpha$, $\beta$, and $passes$, we follow best practices suggested by prior work~\cite{LDApass,Wallach09LDAparam}.

Jaccard similarity measures the similarity between two adjacent topics using Equation~\eqref{eq:jaccard}~\citep{Agrawal, Greene}:  

    \begin{equation}
        J(w_i,w_j) = \frac{(w_i \cap w_j)}{(w_i \cup w_j)} \label{eq:jaccard}
    \end{equation}
    
\noindent where $i$ and $j$ represent different topics, and $w_i$ and $w_j$ are sets of keywords in each topic.
A Jaccard similarity of 1 indicates topics share all keywords, while 0 means they have no common keyword.

We use Jaccard similarity to maximize the topic divergence and minimize the degree of topic overlap, defined by Equation~\eqref{eq:Jaccardi}:
%
    \begin{equation}
        Sim(LDA) = mean(\{ J(w_i,\;w_j) \;\mid\; \forall w_i, w_j \in LDA \}) \label{eq:Jaccardi}
    \end{equation}
\noindent where $LDA$ is the computed LDA model, $w_i$ and $w_j$ represent each pair of topics in the model.
    
Topic coherence measures the degree of semantic similarity of words within each topic~\citep{Newman}. 
We compute topic coherence {\tt Coh(LDA, corpus)} using {\tt Gensim}'s {\tt CoherenceModel}~\citep{Roder}, where {\tt LDA} is the LDA model and {\tt corpus} is the pre-processed texts.

To determine the optimal number of topics, we maximize intra-topic coherence while minimizing inter-topic similarity, as shown in the optimization function in Equation~\eqref{eq:maxJaccard}:
%
%
    \begin{equation}
        \operatorname*{arg\,max}_i f(i) = Coh(LDA(i),\;corpus) - Sim(LDA(i)) \label{eq:maxJaccard}
    \end{equation}
\noindent where $i$ is the number of topics, $Coh$ measures topic coherence between the LDA model with $i$ topics and the corpus, and $Sim(i)$ is the average topic overlap computed by Equation~(\ref{eq:Jaccardi}).

\MyPara{Topic Assignment} 
After performing 100 iterations with 1-100 topics, we determine that the optimal number of topics is 8.
%
To name the topics, we randomly select 235 posts with accepted answers (at most 30 from each topic, ensuring a 97\% confidence level with a 5\% margin of error) and manually summarize the keywords into a high-level description. 
The resulting topics, keywords, and our manually determined descriptive names are shown in Table~\ref{tab:RQ2}.
LDA assigns topic membership probabilities to each post, indicating how likely a post belongs to each topic.  We assign each post to its most probable topic.
For example, \sopost{61882}~\cite{so61882} has the likelihood of 92\% being in topic 7 (\mobile), 7.3\% being in topic 6 (\polling), etc., thus it is assigned to topic 7 (\mobile).

To further understand energy-related posts, we conduct a systematic analysis to each LDA topic, specifically: (1)~computing their popularity; and (2)~calculating difficulty-related metrics that represent how difficult it is to receive a satisfactory answer.

\MyPara{Computing Popularity} 
We measure relative topic popularity using Equation~\eqref{popularity} proposed by~\citet{Pinto}:
\begin{equation}\label{popularity}
\begin{split}
\mathbb{P} & = \mathbb{S} + \mathbb{A} + \mathbb{C} + \mathbb{V}\\
\end{split}
\end{equation}
\noindent
where $\mathbb{S}$ is the question score, $\mathbb{A}$ is the accepted answer score, $\mathbb{C}$ is the number of comments, and $\mathbb{V}$ is the normalized view count.
$\mathbb{V}$ is computed using Equation~\eqref{visualizations}:
%
\begin{equation}\label{visualizations}
\begin{split}
\mathbb{V} & = \frac{\tt QuestionViews}{\tt TotalViews} \\
\end{split}
\end{equation}
\noindent
where $\tt QuestionViews$ is the current question's view count, and $\tt TotalViews$ is the total view count of all questions in our dataset.

Note that although the original computation of popularity in \citet{Pinto} includes the favorization score of questions (i.e., how many users favorized the question), this feature is deprecated on Stack Overflow and has been removed in the latest data dump we obtain.  Thus we exclude it from our computation.

All of $\mathbb{S}$, $\mathbb{A}$, $\mathbb{C}$, and $\mathbb{V}$ are normalized using min-max normalization in Equation~\eqref{eq:minmax}, to \Revision{mitigate the problem that the different metrics can have different scales, or one metric dominates the calculation of popularity}:
\begin{equation}\label{eq:minmax}
    X_{norm} = \frac{x-min(x)}{max(x)-min(x)}
\end{equation}
%
\noindent 
where $x$ is the original value, and $min(x)$ and $max(x)$ are the minimum and maximum values in our dataset.

\MyPara{Calculating Difficulty-Related Metrics}
We calculate a set of metrics to characterize the level of difficulty of the questions related to each topic. 
\Revision{The difficulty metrics include the proportion of questions with an accepted answer and without an answers, as well as the median response time (in hours) for receiving the first answer and the first comment.}

\input{Tables/Topic_popularity}
\input{Tables/Topic_hardness}

\subsubsection{Findings}\label{rq3-findings}

\textbf{We discern eight topics from the energy-related questions.} 
Table~\ref{tab:lda-topic-defin} lists the topics with their definitions, examples, and popularity. The most frequently discussed topics are \gps (27.2\% of questions), \resource (18.2\%), and \mobile (13.8\%), followed by \sensor (10.1\%), \polling (9.2\%), \data (8.1\%), \trans (7.9\%), and \thread (5.4\%). 
\Revision{
The alignment of frequency-based (Table~\ref{tab:lda-topic-defin}) and popularity-based (Table~\ref{tab:RQ2-3}) rankings shows that the topics with the most questions tend to accumulate the most engagement (simply counts posts), whereas popularity considers other metrics, such as views and up-votes, to capture overall importance.
}

\textbf{Questions related to \gps receive the most attention from the SO community}, because the positioning services (such as GPS) directly impact battery life while being essential for many modern applications (e.g., navigation).
According to Table~\ref{tab:RQ2-3}, positioning-related questions rank highest in popularity.
For instance, positioning services have long been recognized for their significant impact on energy consumption~\cite{Zhuang10gps}.
Our manual investigation also reveals the challenges in balancing the need of real-time \Revision{position} data and energy efficiency, e.g., practitioners ask questions about
managing \Revision{position} update frequencies (\sopost{26336225}~\cite{so26336225}, \sopost{29616977}~\cite{so29616977}) or time periods (\sopost{23560949}~\cite{so23560949}), 
and using alternative \Revision{position} tracking sensors to reduce power consumption (\sopost{43596157}~\cite{so43596157}, \sopost{10920904}~\cite{so10920904}).

\textbf{\resource is also of high interest to practitioners} due to their critical role in system performance and efficiency. 
The increasing complexity of managing computing resources, particularly in cloud computing and virtualization environments, drives interest in this topic. Practitioners seek to optimize resource utilization to improve overall performance and energy efficiency, at application level (\sopost{58156084}~\cite{so58156084}), framework level (\sopost{53577434}~\cite{so53577434}), and hardware level (\sopost{28003660}~\cite{so28003660}).
There is also a need to measure or estimate the energy consumption of computing resources (\sopost{44228005}~\cite{so44228005}, \sopost{4485153}~\cite{so4485153}).

\textbf{Questions related to \data are the most challenging to receive satisfactory answers.} 
Table~\ref{tab:RQ2-2} shows that \data questions have the highest non-acceptance ratio (71.1\%) and lowest accepted answer ratio (28.9\%), with a median response time of 9.08 hours. 
In contrast, prior studies show better metrics for other domains: \citet{RS} report a 30\% non-acceptance ratio and 21-minute median response time for mobile development questions, while \citet{Bagherzadeh} report a 60.5\% non-acceptance ratio and 3.3-hour median response time for big data questions. 
Our analysis of sampled posts reveals that this difficulty stems from mismatches between developers' expectations and the underlying complexities of data management in energy-sensitive contexts (e.g., \sopost{1738515}~\cite{so1738515}, \sopost{23617388}~\cite{so23617388}, \sopost{54028927}~\cite{so54028927}, and \sopost{8782922}~\cite{so8782922}),
which involve OS-level abstractions, specialized domain knowledge, and a lack of universal libraries.
\textbf{\polling questions take the longest time to receive responses.} 
Table~\ref{tab:RQ2-2} shows that polling-related questions have the longest median time of 12 hours before receiving an answer. 
Our manual analysis reveals that polling typically involves low-level hardware details such as 
microcontroller registers (\sopost{50064364}~\cite{so50064364}), 
clocks (\sopost{27383269}~\cite{so27383269}), 
and ADC sampling (\sopost{53173447}~\cite{so53173447}).
About 62.1\% of answers require in-depth knowledge of specific MCUs or peripherals, however, only a few community members may have that expertise.
Additionally, broad questions requiring clarification further delay responses (\sopost{28760898}~\cite{so28760898}, \sopost{39804144}~\cite{so39804144}). 
The complexity of polling issues---stemming from low-level hardware interactions, specialized domain knowledge, and lack of standards---leads to prolonged answer delays, lower acceptance rates, and more unresolved posts compared to other topics.

\begin{figure}[t]
    \centering
        \vspace{-3mm}
        \includegraphics[width=\linewidth]{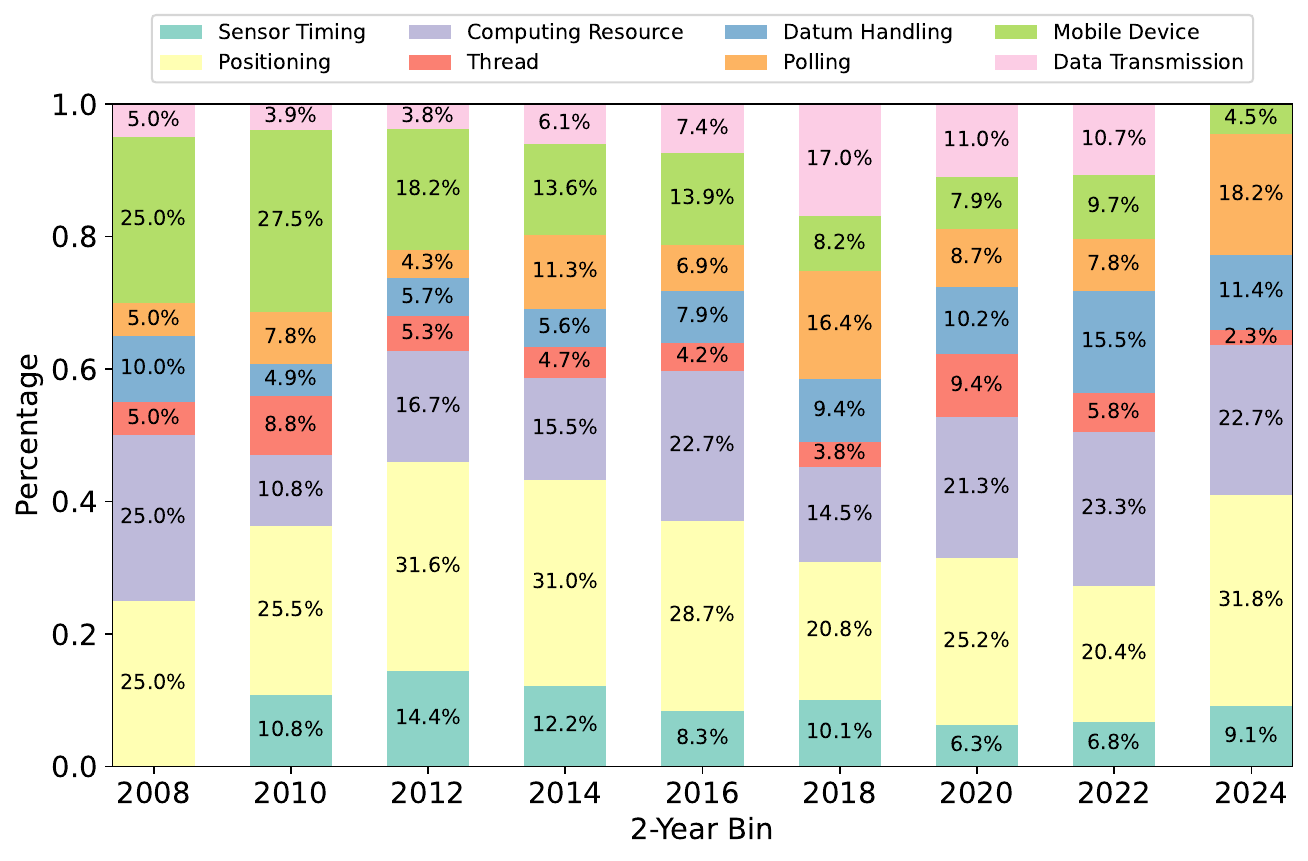}
        \vspace{-8mm}
        \caption{Topic Distribution Every 2 Years.}
        \label{fig:topic-distr}
    \vspace{-5mm}
\end{figure}

\Revision{\textbf{Technological concerns shift from hardware limitations (e.g., \mobile) to data-oriented connective topics (e.g., \trans and \data).}
As shown in Figure~\ref{fig:topic-distr}, 
\trans notably increases from about 5.0\% in 2008 to a peak of 17.0\% in 2018, and stabilizes at around 10-11\%, which reflects growing concerns and developments in big data, cloud computing, IoT, and wireless communication technologies. \data significantly grows, especially from 2016 onwards, indicating increased attention toward data processing, analytics, and big data technologies, which peaks at 15.5\% in 2022, reflecting the rapid adoption of data-intensive applications.
\gps remains consistently significant across all years, indicating its continuous importance in technology, maintaining around 25-32\%. 
\mobile initially dominates but decreases over time, from 25.0\% in 2008 to less than 10\% after 2018, reflecting maturity in mobile technologies or shifting interest towards other emerging technologies. 
}

\Revision{
Pinto et al. identify seven causes of energy-related questions (e.g., ``faulty GPS behaviour'')~\cite{Pinto}. 
Among them, ``unnecessary resource usage'' and ``excessive synchronization'' are related to our topics \resource (throttling idle cores) and \thread (avoiding busy-waits and lock overhead), separately; ``faulty GPS behaviour''  and ``void polling'' can be mapped our topics \gps (minimizing GPS on-time or avoiding fine-grained location) and \polling (favoring interrupt-driven approaches), respectively.
Beyond these overlaps, our study uncovers several new topics (\mobile, \sensor, \data, and \trans) not emphasized by Pinto et al., reflecting how the energy-efficient software landscape has evolved far beyond the CPU and GPS focus of 2013.
Our eight topics not only capture evolving concerns, such as sensor batching and network I/O trade-offs, but also furnish actionable implications, such as 
optimizing \data at the OS/kernel, hardware, and data structure levels to directly control resource usage rather than merely tweaking high‑level application code.
}

\conclusionbox{
\noindent\textbf{\underline{Summary}:} We derive eight common topics using LDA: \gps, \resource, \mobile, \sensor, \polling, \data, and \thread.
Our results indicate that questions related to \gps are the most prevalent, reflecting practitioners' strong interest in optimizing battery performance for location-based services.
Meanwhile, questions about \polling and \data are the most difficult to receive community support due to in-depth knowledge required.
\Revision{Moreover, our topic evolution analysis reveals a shift from hardware-centric concerns toward data-oriented topics.}
Our inspections suggest that providing more context-specific support (e.g., on how to efficiently handle positioning) could help practitioners overcome challenges in energy-efficient software development; our findings highlight these specific areas.
}

%% file: Tables/Topic_def.tex
\begin{table*}
\centering
\caption{Definitions of the derived energy consumption related topics.}
\label{tab:lda-topic-defin}
\vspace{-3mm}
\resizebox{\linewidth}{!}{
\begin{tabular}{|l|l|l|}
\hline
\rowcolor[HTML]{C0C0C0} 
\multicolumn{1}{|c|}{\cellcolor[HTML]{C0C0C0}Topic} & \multicolumn{1}{c|}{\cellcolor[HTML]{C0C0C0}Definition (D) - Quote (Q)}                                                                                                                                                                                                                             & \multicolumn{1}{c|}{\cellcolor[HTML]{C0C0C0}Freq} \\ \hline
Positioning                                                  & \begin{tabular}[c]{@{}l@{}}\textbf{D}: Questions that seek to improve the efficiency of location updates, reduce GPS-related battery drain, \\ and balance real-time tracking with energy savings.\\ \textbf{Q}: ``\textit{How to energy efficiently track GPS/[...]}''~\cite{so10920904}\end{tabular} & 325 (27.2\%)                                      \\ \hline
\rowcolor[HTML]{C0C0C0} 
Computing Resource                                   & \begin{tabular}[c]{@{}l@{}}\textbf{D}: Questions that concern about perceiving the use of computing resources and the efficiency of resource allocation.\\ \textbf{Q}: ``\textit{Is there a way to use processors/[...] more effectively?}''~\cite{so58156084}\end{tabular}                        & 217 (18.2\%)                                      \\ \hline
Mobile Device                                        & \begin{tabular}[c]{@{}l@{}}\textbf{D}: Questions about balancing power states, or device-specific features on mobile devices.\\ \textbf{Q}: ``\textit{How to sense phone's mode/[...] to save power consumption?}"~\cite{so5151872}\end{tabular}                                                      & 165 (13.8\%)                                      \\ \hline
\rowcolor[HTML]{C0C0C0} 
Sensor Timing                                        & \begin{tabular}[c]{@{}l@{}}\textbf{D}: Questions that focus on optimizing the scheduling and continuous operation of sensors within an app.\\ \textbf{Q}: ``\textit{Why [...] every second/minute from a service would consume so much power?}''~\cite{so5765212}\end{tabular}                         & 120 (10.1\%)                                      \\ \hline
Polling                                              & \begin{tabular}[c]{@{}l@{}}\textbf{D}: Questions about concerns about overhead due to continuously ``polling'' hardware or software flags.\\ \textbf{Q}: ``\textit{How to reduce instruction count/[...] in code/clock sources?}''~\cite{so53173447}\end{tabular}                                        & 110 (9.2\%)                                       \\ \hline
\rowcolor[HTML]{C0C0C0} 
Datum Handling                                       & \begin{tabular}[c]{@{}l@{}}\textbf{D}: Questions related to storing, selecting, or allocating data efficiently.\\ \textbf{Q}: ``\textit{Is it optimized in terms of processing/allocating/sending/storing data [...]?}''~\cite{so1738515}\end{tabular}                                                 & 97 (8.1\%)                                        \\ \hline
Data Transmission                                    & \begin{tabular}[c]{@{}l@{}}\textbf{D}: Questions that revolve around capturing, sending, or synchronizing data or messages between devices. \\ \textbf{Q}: ``\textit{How to send data to a server [...] with less energy consumption?}''~\cite{so53359800}\end{tabular}                                & 94 (7.9\%)                                        \\ \hline
\rowcolor[HTML]{C0C0C0} 
Thread                                               & \begin{tabular}[c]{@{}l@{}}\textbf{D}: Questions that involve synchronizing states between threads or processes.\\ \textbf{Q}: ``\textit{How to manage thread/[...] execution efficiently?}''~\cite{so61884014}\end{tabular}                                                                           & 65 (5.4\%)                                        \\ \hline
\end{tabular}
}
\vspace{-4mm}

\end{table*}

%% file: Tables/Topic_popularity.tex
\begin{table}
\caption{Common energy-consumption topics with cumulative normalized popularity score metric. The \textbf{bold} numbers highlight the most popular topic in terms of each metric.}
\label{tab:RQ2-3}
\vspace{-2.5mm}
\centering
\begin{footnotesize}
\begin{threeparttable}
\begin{tabular}{c|c|cccc}
\hline
\rowcolor[HTML]{C0C0C0} 
Topic              & P             & S             & A            & C             & V            \\ \hline
\textbf{Positioning}       & \textbf{61.5} & \textbf{26.5} & \textbf{8.2} & \textbf{20.5} & \textbf{6.4} \\
\rowcolor[HTML]{C0C0C0} 
Computing Resource & 53.3          & 19.9          & 7.5          & 20.3          & 5.6          \\
Mobile Device      & 31.9          & 13.4          & 4.5          & 9.8           & 4.1          \\
\rowcolor[HTML]{C0C0C0} 
Sensor Timing      & 22.1          & 9.4           & 2.4          & 8.5           & 1.8          \\
Polling            & 21.9          & 8.5           & 2.4          & 8.8           & 2.2          \\
\rowcolor[HTML]{C0C0C0} 
Datum Handling     & 23.3          & 7.7           & 2.1          & 10.8          & 2.7          \\
Data Transmission  & 17.4          & 6.7           & {\ul 1.9}    & 7.4           & {\ul 1.3}    \\
\rowcolor[HTML]{C0C0C0} 
{\ul Thread}       & {\ul 13.6}    & {\ul 5.7}     & 2.0          & {\ul 4.4}     & 1.5          \\ \hline
\end{tabular}
\begin{tablenotes}
      \scriptsize
      \item * P represents the popularity of the question, S is the score of the question, A denotes the adopted answer score of the question, and C and V stand for comments and the normalized number of views, respectively. 
\end{tablenotes}
\end{threeparttable}
\end{footnotesize}
\vspace{-3.5mm}
\end{table}

%% file: Tables/Topic_hardness.tex
\begin{table}[t!]
\caption{Common energy-consumption topics with their metrics representing difficulties. The \textbf{bold} numbers highlight the most difficult topic in terms of each metric.}
\label{tab:RQ2-2}
\vspace{-2.5mm}
\centering
\resizebox{\linewidth}{!}{
\begin{tabular}{
>{\columncolor[HTML]{D7D7D7}}c c
>{\columncolor[HTML]{D7D7D7}}c c
>{\columncolor[HTML]{D7D7D7}}c }
\hline
Topic              & \begin{tabular}[c]{@{}c@{}}With an \\accepted answer\end{tabular} & \begin{tabular}[c]{@{}c@{}}Without an\\ answer\end{tabular} & \begin{tabular}[c]{@{}c@{}} Receiving an \\answer (Hr)\end{tabular} & \begin{tabular}[c]{@{}c@{}}Receiving a \\comment (Hr)\end{tabular} \\ \hline
Positioning                & 36.0\%                                                                                     & 21.8\%                                                                               & 2.25                                                                                & 0.32                                                                                \\
Computing Resource & 39.6\%                                                                                     & 18.9\%                                                                               & 3.42                                                                                & 0.27                                                                                \\
Mobile Device      & 37.6\%                                                                                     & 20.6\%                                                                               & 1.48                                                                                & 0.39                                                                                \\
Sensor Timing      & 38.3\%                                                                                     & 25.0\%                                                                               & 1.75                                                                                & 0.22                                                                                \\
Polling            & 33.6\%                                                                                     & 26.4\%                                                                               & \textbf{12.0}                                                                      & 0.63                                                                                \\
Datum Handling     & \textbf{28.9\%}                                                                            & \textbf{29.9\%}                                                                      & 9.08                                                                                & 0.46                                                                                \\
Data Transmission  & 37.2\%                                                                                     & 19.1\%                                                                               & 3.12                                                                                & 0.33                                                                                \\
Thread             & 53.8\%                                                                                     & 23.1\%                                                                               & 1.36                                                                                & \textbf{0.78}                                                                       \\ \hline
Average            & 38.1\%                                                                                     & 23.1\%                                                                               & 4.31                                                                                & 0.42                                                                                \\ \hline
\end{tabular}
}
\vspace{-4mm}
\end{table}

%% file: Text/RQ2.tex
\subsection{RQ2: What are the intentions behind the energy-efficient development questions?}\label{rq2}

\subsubsection{Motivation}

Practitioners seek assistance on SO with various intentions, from understanding concepts to troubleshooting API usages. 
The tags associated with the questions usually do not reveal why questions are asked. 
By categorizing these intentions across different topics (identified in RQ1), we can uncover the most common challenges in energy-efficient software development and provide deeper insights into practitioners' challenges and needs.

\subsubsection{Approach}
We categorize question intentions based on the taxonomy proposed by \citet{category}, which includes seven categories: {\tt API USAGE}, {\tt CONCEPTUAL}, {\tt DISCREPANCY}, {\tt ERRORS}, {\tt REVIEW}, {\tt API CHANGE}, and {\tt LEARNING}. 
We employ three state-of-the-art LLMs to classify question intentions: QwQ-32b (zero-shot)~\cite{qwq32b}, GPT-4o (zero-shot)~\cite{4o}, and Deepseek-R1 (one-shot)~\cite{deepseekr1}. 
We follow best practices in prompt engineering~\cite{openai-prompt, huggingface-prompt} and prior work~\cite{fathollahzadeh25towards, triem24tipping, khan24debating} to optimize model performance, and select zero-shot/one-shot prompting style based on preliminary experiments.
In our experiments, the scenario where all three LLMs produce different intents does not occur.
We statistically sample a subset (277 posts, sufficient for covering a 90\% confidence level and 5\% interval) and manually verify the LLM-voted results. 
\Revision{At least two LLMs produce the same intent in the majority voting.
We compare the manually labeled intents with LLM-voted results (determined by the majority) with a Cohen's Kappa of 0.64, indicating substantial consensus between human work and LLMs.}

\input{Tables/Intentions}
\input{Tables/Topic_w_intents}

\subsubsection{Findings}
\textbf{Nearly half of energy-related questions are \concept, showing that understanding energy fundamentals is a primary concern.}
Table~\ref{tab:RQ1} lists the number of questions for each intention category.
Understanding concepts (\concept) is the most common intention, accounting for 48.6\% of questions.
This is in stark contrast to the prior work~\citep{category} studying Android development SO questions, where only 26.80\% of the questions pertain to \concept. 
%
Through in-depth analyses, we identify two reasons for this high proportion of conceptual questions:
(1)~energy consumption spans diverse hardware and software platforms, requiring different measurement approaches and tools. 
For example, \sopost{66697291}~\cite{so66697291} shows that measuring energy consumption of Python scripts requires platform-specific hardware, software, and APIs;
and (2)~as confirmed by prior work~\cite{Pang, Manotas}, developers struggle with fundamental energy concepts, API limitations, and behavior. 
For example, \sopost{48766582}~\cite{so48766582} illustrates how developers working on battery-powered ARM-based embedded Linux systems face challenges in optimizing power consumption and selecting appropriate kernel APIs. 



\textbf{\concept issues dominate energy-related discussions on multiple topics, especially for the \mobile topic, implying the complexity of balancing power states on mobile devices.} 
Table~\ref{tab:topic_w_intents} shows the intention distribution across topics, where the \mobile topic has the highest proportion (59.4\%) of \concept issues.
Our manual analysis reveals three key reasons: 
(1)~Developers must reconcile diverse OS power models, requiring foundational concepts like wake locks and battery monitoring before implementation (\sopost{724349}~\cite{so724349}); 
(2)~Heterogeneous hardware features across devices (e.g., AMOLED screens, custom power-saving modes) necessitate understanding how components interact to affect battery life (\sopost{2902382}~\cite{so2902382}); 
and (3)~Developers need to grasp the principles of CPU usage, screen brightness, sensor polling, and network connectivity to avoid degrading user experience or battery life (\sopost{38635068}~\cite{so38635068}). 

\textbf{Approximately one-third of the energy-related questions concern {\tt API USAGE}, reflecting developers' need for practical implementation guidance.}
{\tt API USAGE} questions account for 37.0\% of energy-related SO questions, similar to the 38.8\% observed in Android development SO questions~\cite{category}. 
These questions often focus on the practices of using certain APIs in energy-efficient ways, such as \sopost{38158828}~\cite{so38158828} and \sopost{9863131}~\cite{so9863131} that ask about location services. 
As applications for energy-constrained platforms proliferate, developers increasingly need to optimize system resources (e.g., CPU, memory, and battery) while navigating compatibility challenges across diverse APIs. 

\textbf{\thread management is a particularly challenging area, as evidenced by the high frequency of \api questions within this topic.}
Table~\ref{tab:topic_w_intents} reveals that \thread-related questions have the highest proportion (44.6\%) of \api issues. 
Our investigation identifies three key challenges: 
(1)~threads require precise control for starting, stopping, and synchronization, with improper management leading to energy-wasting issues like infinite loops or race conditions (\sopost{61884014}~\cite{so61884014}); 
(2)~efficiently waking threads from idle states is difficult, as poor timing can cause delays that impact responsiveness (\sopost{51973350}~\cite{so51973350}); 
and (3)~developers struggle to choose between continuous event monitoring and periodic activation (e.g., using alarm services) to balance power consumption and performance (\sopost{15698999}~\cite{so15698999}). 

\textbf{Practitioners face challenges in resolving discrepancies in energy-aware development.}
{\tt DISCREPANCY} issues rank third (22.8\%) among energy-related questions, reflecting the need to troubleshoot energy consumption bugs. 
In \sopost{42850419}~\cite{so42850419}, a developer seeks help resolving unexpected power consumption in their app and is unsure which components may be relevant to the issue. 
Energy-related discrepancies are particularly challenging because they require understanding complex interactions between system components and identifying inefficiencies that may not be immediately apparent. 
The abstract nature of energy consumption, leading to fewer resources and solutions available for practitioners to refer to, further complicates troubleshooting.

\textbf{\data-related issues are particularly challenging in energy-efficient development, reflecting the intricate nature of efficiently handling data under energy constraints.}
We observe that \data-related questions have the highest frequency (36.1\%) of discrepancy questions.
Our investigation reveals three key challenges: 
(1)~understanding data allocation's impact on power consumption requires deep system-level insights that many developers lack (\sopost{1738515}~\cite{so1738515}); 
(2)~inefficient data processing patterns (looping, filtering, selecting) can cause unexpected energy consumption that is difficult to diagnose (\sopost{23617388}~\cite{so23617388}); 
and (3)~retrieving data from external systems or libraries often leads to misalignments between expected and actual data definitions, causing energy inefficiencies (\sopost{54028927}~\cite{so54028927}). 
Even minor errors in data processing can significantly impact energy consumption, making data handling particularly challenging in energy-constrained environments. 
This complexity highlights the need for better tools and guidelines to help developers implement energy-efficient data processing patterns.

\begin{figure}
    \centering
        \includegraphics[width=1.01\linewidth]{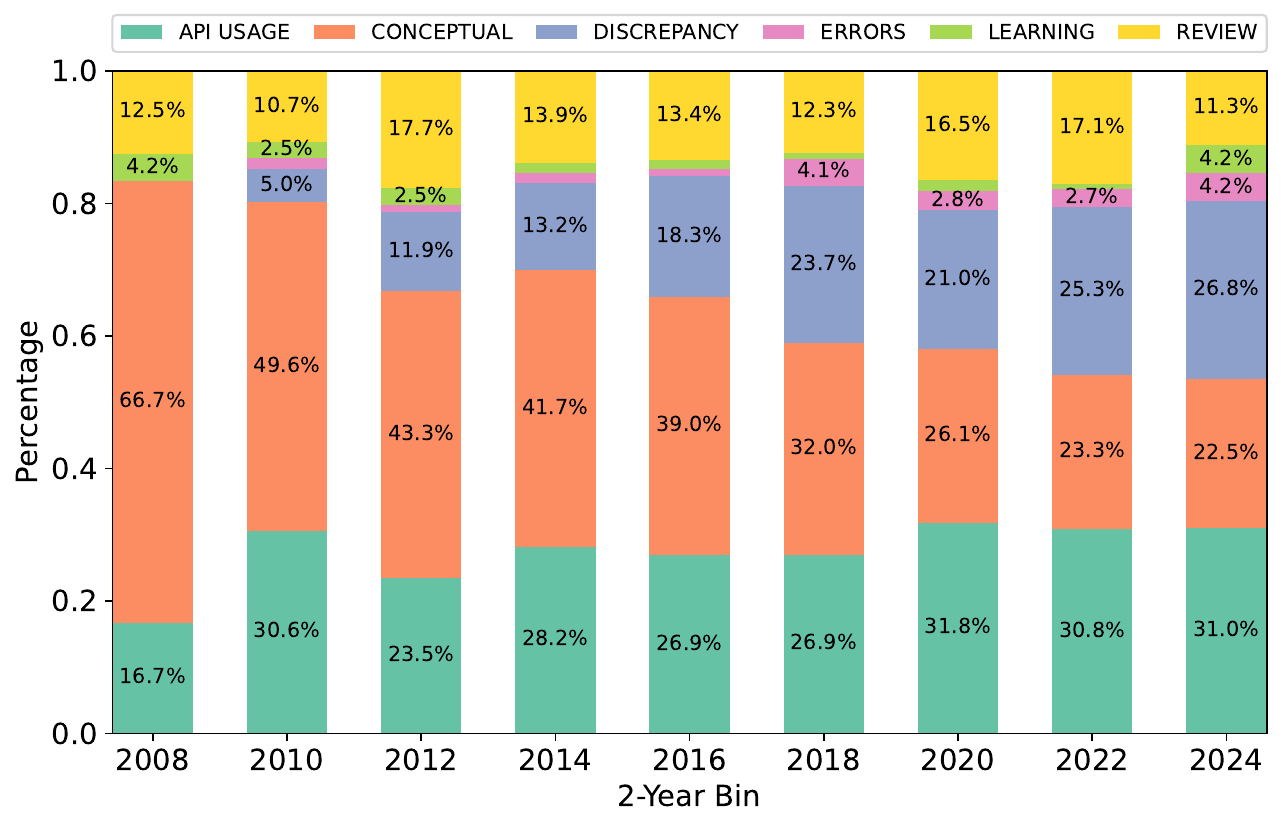}
        \vspace{-8mm}
        \caption{Intent Distribution Every 2 Years.}
        \label{fig:intent-distr}
    \vspace{-6mm}
\end{figure}

\Revision{\textbf{Intentions of developing energy-efficient software shift from primarily concerning \concept issues toward addressing unexpected results/behavior problems (\disc) over the years}, highlighting evolving focus from theoretical exploration to practical unexpected results/behavior-handling in energy-efficient coding.
Pinto et al. identify the theme {\it General Knowledge} (18.3\% by 2013), which is related to \concept in our study. 
As shown in Figure \ref{fig:intent-distr}, \concept questions have been declining over the years, which indicates improvements in the availability and clarity of foundational resources and conceptual documentation for energy-efficient programming.
However, the proportion of \disc issues rises dramatically from 5.0\% in 2010 to around 27\% by 2024, reflecting a growing demand for actionable guidance on diagnosing and resolving unexpected results/behavior
that impact energy consumption. 
The {\it Code Design} theme in Pinto et al. (questions about programming techniques that can help in saving energy) accounts for 16.5\% of the posts by 2013; 
this theme approximately corresponds to the \review intent in our study, whose frequency is similar (10.7--17.7\%).
}

\conclusionbox{
\noindent\textbf{\underline{Summary}:} 
Nearly half of energy-related questions focus on understanding fundamental concepts, highlighting the complexity of energy consumption as a cross-cutting concern spanning multiple components.
In particular, conceptual questions are particularly prevalent for mobile devices (59.4\%), where developers struggle with diverse operating systems, hardware features, and power management models.
Questions about API usages (37.0\%) and discrepancies (22.8\%) are also common, reflecting the need for practical guidance and troubleshooting resources. 
\Revision{Our intent evolution analysis reveals a shift from theoretical exploration (\concept) to practical unexpected results/behavior-handling (\disc).}
}

%% file: Tables/Intentions.tex
\begin{table}
\centering
\caption{Number of posts per respective intention category.}
\label{tab:RQ1}
\vspace{-3mm}
\begin{threeparttable}
\resizebox{\linewidth}{!}{
\begin{tabular}{c|llc}
\hline
\rowcolor[HTML]{C0C0C0} 
Intention Category & \multicolumn{1}{c}{\cellcolor[HTML]{C0C0C0}Definition}                                                                                                                                                          & \multicolumn{1}{c}{\cellcolor[HTML]{C0C0C0}Example} & Freq                                                    \\ \hline
Conceptual         & \begin{tabular}[c]{@{}l@{}}Questions that understand concepts and ask\\ the limitations of an API and API behavior\end{tabular}                                                                                 & 30027148                                            & \begin{tabular}[c]{@{}c@{}}586 \\ (48.6\%)\end{tabular} \\
\rowcolor[HTML]{C0C0C0} 
API Usage          & \begin{tabular}[c]{@{}l@{}}Questions about how to implement certain\\ functionality or how to use an API\end{tabular}                                                                                           & 3412026                                             & \begin{tabular}[c]{@{}c@{}}441\\ (37.0\%)\end{tabular}  \\
Discrepancy        & \begin{tabular}[c]{@{}l@{}}Questions related to exception problems that the\\ observed result is different from the expectation\end{tabular}                                                                    & 22339063                                            & \begin{tabular}[c]{@{}c@{}}272\\ (22.8\%)\end{tabular}  \\
\rowcolor[HTML]{C0C0C0} 
Review             & \begin{tabular}[c]{@{}l@{}}Questions that ask for the best practice\\ approaches or ask for help to make decisions\end{tabular}                                                                                 & 6866236                                             & \begin{tabular}[c]{@{}c@{}}230\\ (19.3\%)\end{tabular}  \\
Errors             & \begin{tabular}[c]{@{}l@{}}Questions about problems of exceptions with\\ or without code snippets, as well as requiring\\ help in fixing the error or understanding the\\ meaning of the exception\end{tabular} & 60771095                                            & \begin{tabular}[c]{@{}c@{}}33\\ (2.8\%)\end{tabular}    \\
\rowcolor[HTML]{C0C0C0} 
Learning           & \begin{tabular}[c]{@{}l@{}}Questions that ask for documentation or tutorials\\ to learn a tool or language by their own, without \\ asking for a specific instruction or solution\end{tabular}                  & 63105570                                            & \begin{tabular}[c]{@{}c@{}}28\\ (2.3\%)\end{tabular}    \\ \hline
\end{tabular}
}
\begin{tablenotes}
      \scriptsize
      \item * Posts are visitable at \url{https://stackoverflow.com/questions/{Example ID}}
\end{tablenotes}
\end{threeparttable}
\vspace{-3mm}
\end{table}

%% file: Tables/Topic_w_intents.tex
\begin{table}[]
\caption{Intents behind each topic.}
\label{tab:topic_w_intents}
\vspace{-3mm}
\resizebox{\linewidth}{!}{
\begin{threeparttable}
\begin{tabular}{l
>{\columncolor[HTML]{D7D7D7}}c c
>{\columncolor[HTML]{D7D7D7}}c c
>{\columncolor[HTML]{D7D7D7}}c c}
\hline
                   & CONCEPTUAL      & API USAGE       & DISCREPANCY     & REVIEW          & ERRORS         & LEARNING        \\ \hline
Positioning                & 50.8\%          & 40.6\%          & 16.9\%          & 23.1\%          & 1.5\%          & 1.5\%           \\
Computing Resource & 54.4\%          & 27.6\%          & 24.0\%          & 16.1\%          & 1.4\%          & \textbf{3.22\%} \\
Mobile Device      & \textbf{59.4\%} & 39.4\%          & 8.5\%           & 9.1\%           & 1.8\%          & 3.0\%           \\
Sensor Timing      & 36.7\%          & 43.3\%          & 25.0\%          & \textbf{29.2\%} & 3.3\%          & 1.7\%           \\
Polling            & 50.0\%          & 34.5\%          & 30.9\%          & 10.9\%          & 4.5\%          & 1.8\%           \\
Datum Handling     & 43.3\%          & 34.0\%          & \textbf{36.1\%} & 22.7\%          & 5.2\%          & 2.1\%           \\
Data Transmission  & 39.4\%          & 34.0\%          & 34.0\%          & 22.3\%          & \textbf{5.3\%} & 3.19\%          \\
Thread             & 41.5\%          & \textbf{44.6\%} & 30.8\%          & 23.1\%          & 4.6\%          & 3.1\%           \\ \hline
\end{tabular}
\begin{tablenotes}
      \item * Each cell represents the frequency of such intent relative to the total number of posts within the corresponding topic. The sum of the frequency values can be greater than 100\% since one post may have more than one intentions.
\end{tablenotes}
\end{threeparttable}
}
\vspace{-6mm}
\end{table}

%% file: Text/RQ3.tex
\noindent\textbf{\subsection{RQ3: What technologies are concerned in energy-efficient development?} \label{rq3}}

\subsubsection{Motivation}
Stack Overflow questions are usually tagged with their relevant technologies, e.g., operating systems (Linux, Windows, etc.), programming languages (Java, C, etc.), and hardware components (mobile, server, etc.). 
Identifying which technologies are most frequently discussed helps direct technical support toward addressing the most prevalent energy-related challenges faced by practitioners. 

\subsubsection{Approach}
We infer concerned technologies from tags by following steps:

\input{Tables/Tag_category}
\input{Tables/tag_specifications}

\MyPara{Merging Synonymous Tags}  
The same technology may sometimes be described with several semantically similar tags, e.g., ``bluetooth'' and ``bluetooth-lowenergy'', which we call synonymous tags.
We merge synonymous tags into a single category, specifically the lower-frequency tag is merged into the higher-frequency tag (e.g., ``bluetooth-lowenergy'' is merged into ``bluetooth'').

\MyPara{Selecting Frequently Occurring Tags} 
After merging synonymous tags, we have 1,008 unique tags occurring 4,803 times in total.
The tags follow a long-tail distribution, where many tags only appear once or twice in our dataset.
To focus on the most important technologies, we follow the Pareto principle (i.e., 80/20 rule) and select the top frequently occurring tags that account for at least 80\% of all occurrences.
After including tags with the same frequency, we have 363 tags covering 84.2\% of all tag occurrences.
We further inspect remaining tags and decide to focus on the tags related to software development technologies (77 out of 363), as the remaining tags are irrelevant or infrequent (many with $\leq$1\% frequency) and may introduce noises to the analysis.

\MyPara{Grouping Tags}
We group the 77 tags into three main categories: operating systems (\os, 10 tags, e.g., android, ios, and linux), programming languages (\pl, 23 tags, e.g., java, c, and python), and hardware (\hw, 44 tags).
The hardware category is further divided into subcategories of mobile (e.g., ipad), accessory (e.g., bluetooth), server (e.g., cloud), embedded (e.g., rasberry-pi), and processing units (e.g., cpu, gpu, and memory).
\Revision{Among the posts, 325 (27.2\%) include multiple tags that span different categories.
We experimented with making tags unique per post by randomly selecting one of multiple tags, which led to a frequency shift by only -1.7\% to 2.3\% per category from our results. 
We choose to keep multiple tags, as they are assigned by SO users in arbitrary orders and there is no evidence to indicate the importance of tags.}

\MyPara{Counting Posts}
For each tag, we count the number of posts associated with it to determine its prevalence in energy-related discussions \Revision{with regard to topics (\S\ref{rq1}) and developer intents (\S\ref{rq2})}. 
This helps us identify which technologies are most frequently concerned by practitioners in relation to energy consumption.


\subsubsection{Findings}

\textbf{\os is the most discussed category of technologies in energy-related posts.}
Table~\ref{tab:R3-tag} lists the technologies and their corresponding tags, with the numbers of posts belong to each technology category. 
\os-related tags appear in 533 posts, more than \hw (415 posts) and \pl (357 posts). 
Note that posts can have multiple tags (e.g., \sopost{17134522}~\cite{so17134522} has both embedded and Objective-C tags), the percentages across categories do not sum to 100\%. 

\Revision{\textbf{A higher number of energy-related questions are found in the context of Mobile OS than PC/server OS.}}
Table~\ref{tab:R3-lda-os} shows the percentage of posts for each specific technology, with Android (28.1\%) and iOS (11.5\%) together accounting for nearly 40\% of all posts.
Meanwhile, Linux (6.2\%), Windows (4.9\%), and MacOS (1.3\%) collectively represent only about 12\%. 
This aligns with \citet{Bao16Android}'s findings on the complexity of Android power management, where different app categories require distinct power management strategies.

\textbf{Resolving energy-related questions often requires referring to other knowledge sources via links.}
Our manual inspection reveals that over 60\% the accepted answers require linking to external knowledge, such as documentations of APIs and tools, practical experience, and in-depth knowledge of the underlying concepts
(e.g., \sopost{11366329}~\cite{so11366329}, \sopost{1298600}~\cite{so1298600}, \sopost{3655806}~\cite{so3655806}, \sopost{6051807}~\cite{so6051807}, and \sopost{6553595}~\cite{so6553595}).
Meanwhile, 43.0\% of questions can be resolved using internal resources like other users' solutions, API documentation, software specifications, and debugging 
(e.g., \sopost{13873570}~\cite{so13873570}, \sopost{40090627}~\cite{so40090627}, \sopost{31368757}~\cite{so31368757}, \sopost{8100506}~\cite{so8100506}, and \sopost{11398732}~\cite{so11398732}). 
However, even though APIs provide specifications and documentation, practitioners still face challenges in finding and understanding them, reflected in the fact that 54.0\% of accepted answers include links to supplementary materials to support practitioners better follow and understand the solution and usage.
Additionally, we notice that questions with respect to 
PC/server-based systems tend to focus more on measuring and profiling energy consumption rather than managing hardware complexity 
(e.g., \sopost{49202065}~\cite{so49202065}, \sopost{7312597}~\cite{so7312597}, \sopost{410122}~\cite{so410122}, \sopost{3655806}~\cite{so3655806}, and \sopost{72711521}~\cite{so72711521}).

\begin{figure}
    \centering
    \includegraphics[width=1\linewidth]{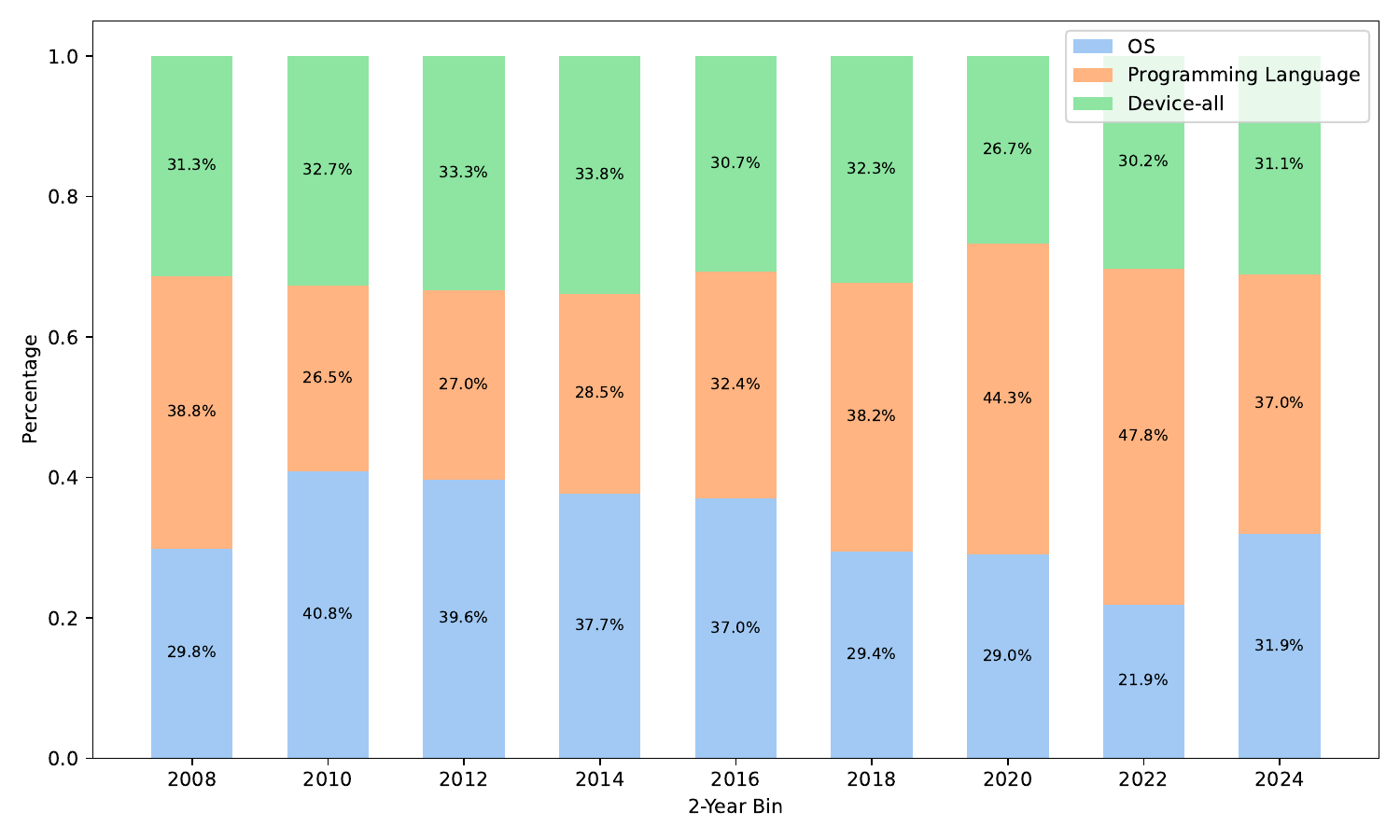}
    \vspace{-8mm}
    \caption{Tag Category Distribution Every 2 Years}
    \label{fig:tag-distr}
    \includegraphics[width=1\linewidth]{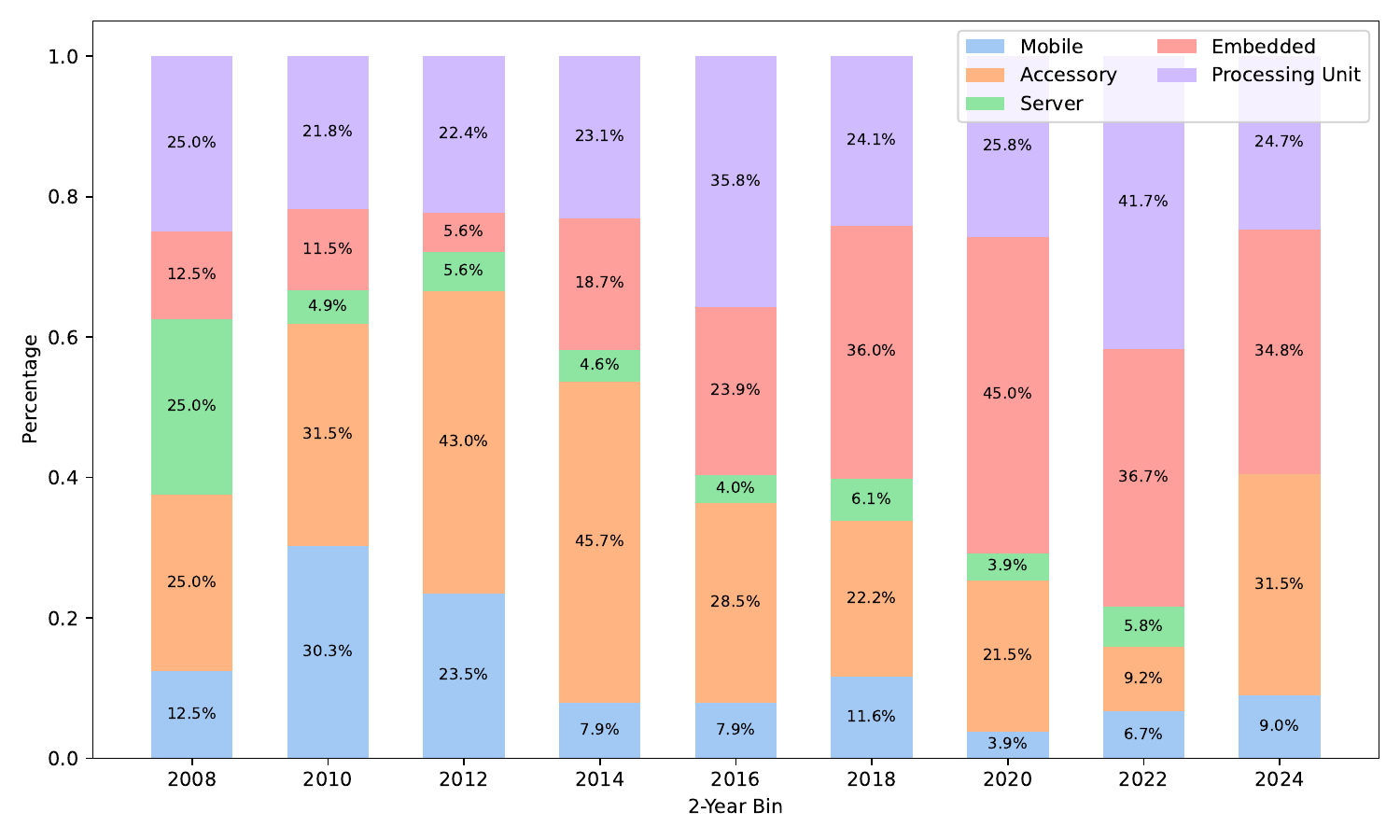}
    \vspace{-8mm}
    \caption{Device Details}
    \label{fig:device-distr}
    \vspace{-6mm}
\end{figure}

\textbf{\os-related energy questions show a declining trend over time.}
Figure~\ref{fig:tag-distr} shows a declining trend in \os-related energy questions over time\footnote{\Revision{The increasing trend from 2022 to 2024 is mainly due to the small sample size in 2024.}}. This likely reflects three developments: 
(1)~cloud providers now handle energy optimization at the infrastructure level; 
and (2)~modern OS kernels have integrated sophisticated power-saving mechanisms like CPU scheduling and dynamic resource allocation~\cite{scheduling, macKernal, Siemers25OS}.





\textbf{Among \hw technologies, practitioners are most concerned about the energy consumption of accessories (e.g., Bluetooth or GPS), followed by processing units (e.g., CPU, GPU, or microchip).}
Table~\ref{tab:R3-lda-device} shows that accessories (14.7\%) are the most discussed hardware components, followed by embedded systems (7.7\%) and CPUs (6.5\%).
Accessories like Bluetooth and GPS draw significant attention because they continuously send and receive signals, consuming substantial energy—a finding consistent with prior work~\cite{Bangash21Location}. 
\Revision{Although modern processors have become energy-efficient through architectural improvements and low-power states like Intel's SpeedStep and AMD's Cool'n'Quiet~\cite{speedstep,quiet},}
processing units (CPU, GPU, and microchips) still receive considerable attention, as they are responsible for executing code and performing calculations.
Figure~\ref{fig:device-distr} reveals interesting trends in hardware concerns: server and accessory-related questions have decreased over time, while embedded hardware questions have increased. 
Server energy concerns have diminished due to cloud migration and mature data center optimizations~\cite{marinescu2022cloud,Zelkowitz11Cloud}. 
Accessory concerns have decreased as components now include robust power-saving features and established best practices~\cite{gomez12BLEoverview}. 
Conversely, embedded hardware questions have surged with IoT growth, increasing microcontroller complexity, and battery-powered device proliferation~\cite{Dragoni17Microservices,Aouedi24IoT}.


\textbf{{\tt Java} is the most concerned single programming language, and the C family of programming languages adding together holds a significant part of the practitioners' concerns.} 
Java is the most discussed individual programming language (5.9\%), likely due to its use in Android development. 
However, the C family of programming languages collectively dominates energy-related discussions, with C (4.9\%), C++ (3.9\%), C\# (2.5\%), and Objective-C (1.4\%) together accounting for 11.3\% of programming language discussions. 
This reflects the widespread use of C family in hardware, OS development (e.g., Linux kernel), and embedded systems.
Figure~\ref{fig:tag-distr} shows increasing attention to programming languages in energy-efficient development. 
This trend emerges as large-scale applications (cloud services and machine learning) written in high-level languages like Java and Python become more prevalent. 
Even small code optimizations can yield substantial energy savings when scaled across thousands of servers~\cite{Chen11Map,benini2011compilers}. 
Additionally, as OS-level power management matures, developers increasingly seek efficiency gains at the application and language levels~\cite{Carroll10phone,Zeng02OS}.



\conclusionbox{
\noindent\textbf{\underline{Summary}:} 
Our analysis of the technologies concerned in energy-related posts highlights the need for targeted support in three key areas: 
(1)~mobile OS energy optimization, where battery constraints and complex component management create unique challenges; 
(2)~efficient hardware utilization, particularly for accessories and embedded systems in IoT applications; 
and (3)~language-specific optimization techniques, especially as high-level languages become more prevalent in large-scale applications. 
The shifting trends suggest that energy optimization focus is moving from system-level to application-level and specialized device contexts.
}

%% file: Tables/Tag_category.tex
\begin{table*}
\caption{Technology specification.}
\label{tab:R3-tag}
\vspace{-3mm}
\centering
\resizebox{\linewidth}{!}{
\begin{threeparttable}
\begin{tabular}{|lll|l|}
\hline
\rowcolor[HTML]{C0C0C0} 
\multicolumn{3}{|c|}{\cellcolor[HTML]{C0C0C0}Tag Category}                                                                                                    & \multicolumn{1}{c|}{\cellcolor[HTML]{C0C0C0}Tag List}                                                                                                                                                                        \\ \hline
\multicolumn{3}{|l|}{OS (533)}                                                                                                                                & android, ios (ios, ios7, and ios8), linux, windows, macos, ubuntu, wear-os, contiki                                                                                                                                          \\ \hline
\rowcolor[HTML]{C0C0C0} 
\multicolumn{3}{|l|}{\cellcolor[HTML]{C0C0C0}Programming Language (357)}                                                                                      & \begin{tabular}[c]{@{}l@{}}java, c, python, c++, swift (swift and swift3), c\#, objective-c, sql, php, asp.net, \\ verilog, node.js, assembly, shell, vhdl, matlab, bash, kotlin, json, jquery, css, javascript\end{tabular} \\ \hline
\multicolumn{1}{|l|}{}                                 & \multicolumn{2}{l|}{Mobile (22)}                                                                     & apple, ipad, mobile, phone, tablet                                                                                                                                                                                           \\ \cline{2-4} 
\multicolumn{1}{|l|}{}                                 & \multicolumn{2}{l|}{\cellcolor[HTML]{C0C0C0}Accessory (175)}                                         & \cellcolor[HTML]{C0C0C0}usb, battery, accelerometer, beacon, bluetooth, gps, ibeacon, usb, wifi, sensors                                                                                                                     \\ \cline{2-4} 
\multicolumn{1}{|l|}{}                                 & \multicolumn{2}{l|}{Server (7)}                                                                      & server, computer, cloud                                                                                                                                                                                                      \\ \cline{2-4} 
\multicolumn{1}{|l|}{}                                 & \multicolumn{2}{l|}{\cellcolor[HTML]{C0C0C0}Embedded (92)}                                           & \cellcolor[HTML]{C0C0C0}embedded, stm32, esp8266, esp32, arduino, pi, raspberry, fpga, xilinx, iot                                                                                                                           \\ \cline{2-4} 
\multicolumn{1}{|l|}{}                                 & \multicolumn{1}{l|}{}                                        & Central (77)                          & microchip, microcontroller, amd, arm, core, cpu, intel, x86, msr                                                                                                                                                             \\ \cline{3-4} 
\multicolumn{1}{|l|}{}                                 & \multicolumn{1}{l|}{}                                        & \cellcolor[HTML]{C0C0C0}Graphics (39) & \cellcolor[HTML]{C0C0C0}gpgpu, gpu, nvidia, cuda                                                                                                                                                                             \\ \cline{3-4} 
\multicolumn{1}{|l|}{\multirow{-7}{*}{Hardware (415)}} & \multicolumn{1}{l|}{\multirow{-3}{*}{Processing Unit (123)}} & Storage (12)                          & memory                                                                                                                                                                                                                       \\ \hline
\end{tabular}
\begin{tablenotes}
      \scriptsize
      \item * The number of posts for each tag category or sub-category is listed in parentheses (e.g., OS (533)).\\
      \item ** Since a post may span multiple tag categories,
      the sum of subcategories does not add up directly (e.g., the number of posts in Processing Unit does not equal to the sum of Central, Graphics, Storage, and Embedded).
\end{tablenotes}
\end{threeparttable}
}
\vspace{-4mm}
\end{table*}

%% file: Tables/tag_specifications.tex
\begin{table*}[]
\caption{Distribution of posts related to technologies.}
\vspace{-3mm}
\label{tab:R3-cat-os}
\label{tab:R3-lda-os}
\label{tab:R3-lda-device}
\label{tab:R3-cat-device}
\label{tab:R3-lda-pl}
\label{tab:R3-cat-pl}
\resizebox{\linewidth}{!}{
\begin{threeparttable}
\begin{tabular}{l|
>{\columncolor[HTML]{D7D7D7}}c c
>{\columncolor[HTML]{D7D7D7}}c c
>{\columncolor[HTML]{D7D7D7}}c |ccccccccc|
>{\columncolor[HTML]{D7D7D7}}c c
>{\columncolor[HTML]{D7D7D7}}c c
>{\columncolor[HTML]{D7D7D7}}c c
>{\columncolor[HTML]{D7D7D7}}c |}
\cline{2-22}
                                            & \multicolumn{5}{c|}{\cellcolor[HTML]{D7D7D7}Operating System}                                                                                                                                                      & \multicolumn{9}{c|}{Programming Language}                                                                                                                                                                                                                                                                                                         & \multicolumn{7}{c|}{\cellcolor[HTML]{D7D7D7}Hardware}                                                                                                                                                                                                       \\ \hline
\multicolumn{1}{|l|}{}                      & \cellcolor[HTML]{D7D7D7}                                   &                       & \cellcolor[HTML]{D7D7D7}                        &                           & \cellcolor[HTML]{D7D7D7}                        &                                 & \cellcolor[HTML]{D7D7D7}                         &                     & \cellcolor[HTML]{D7D7D7}                      &                         & \cellcolor[HTML]{D7D7D7}                      &                       & \cellcolor[HTML]{D7D7D7}                             &                               & \cellcolor[HTML]{D7D7D7}                                     &                            & \cellcolor[HTML]{D7D7D7}                         & \multicolumn{1}{c|}{}                         & \multicolumn{3}{c|}{\cellcolor[HTML]{D7D7D7}Processing Unit} \\ \cline{20-22} 
\multicolumn{1}{|l|}{\multirow{-2}{*}{Tag}} & \multirow{-2}{*}{\cellcolor[HTML]{D7D7D7}\textbf{Android}} & \multirow{-2}{*}{iOS} & \multirow{-2}{*}{\cellcolor[HTML]{D7D7D7}Linux} & \multirow{-2}{*}{Windows} & \multirow{-2}{*}{\cellcolor[HTML]{D7D7D7}MacOs} & \multirow{-2}{*}{\textbf{Java}} & \multirow{-2}{*}{\cellcolor[HTML]{D7D7D7}Python} & \multirow{-2}{*}{C} & \multirow{-2}{*}{\cellcolor[HTML]{D7D7D7}Swift} & \multirow{-2}{*}{C++} & \multirow{-2}{*}{\cellcolor[HTML]{D7D7D7}C\#} & \multirow{-2}{*}{SQL} & \multirow{-2}{*}{\cellcolor[HTML]{D7D7D7}Javascript} & \multirow{-2}{*}{Objective-C} & \multirow{-2}{*}{\cellcolor[HTML]{D7D7D7}\textbf{Accessory}} & \multirow{-2}{*}{Embedded} & \multirow{-2}{*}{\cellcolor[HTML]{D7D7D7}Mobile} & \multicolumn{1}{c|}{\multirow{-2}{*}{Server}} & \textbf{Central}         & Graphics         & Storage        \\ \hline
\multicolumn{1}{|l|}{Freq}                  & \textbf{28.1}\%*                                           & 11.5\%                & 6.2\%                                           & 4.9\%                     & 1.3\%                                           & \textbf{5.9\%}                  & \cellcolor[HTML]{D7D7D7}5.5\%                    & 4.9\%               & \cellcolor[HTML]{D7D7D7}4.0\%                 & 3.9\%                   & \cellcolor[HTML]{D7D7D7}2.5\%                 & 1.8\%                 & \cellcolor[HTML]{D7D7D7}1.6\%                        & 1.4\%                         & \textbf{14.7\%}                                              & 7.7\%                      & 1.8\%                                            & \multicolumn{1}{c|}{0.6\%}                    & \textbf{6.5\%}           & 3.3\%            & 1.0\%          \\ \hline
\end{tabular}
\begin{tablenotes}
      \small
      \item * The figure indicates the percentage of tag category-related posts (e.g., Android) among all posts.
\end{tablenotes}
\end{threeparttable}
}
\vspace{-4mm}
\end{table*}

%% file: Text/Impli.tex
\section{Implications}\label{sec:Implications}


Our findings suggest 
\textbf{teaching and supporting energy-aware development systematically, from \hw (e.g., optimizing the usage of energy-intensive devices such as GPS) to \os (the considerations for servers and mobile operating systems) to \pl (the particularity of different languages' support for energy-aware development).}

Below, we discuss some detailed implications based on our findings.


\Revision{\textbf{Minimizing direct usage of hardware accessories when software-based saving is possible (e.g., by relying on deferred location APIs, or batching/deferring hardware accesses).}}
\S\ref{rq1} emphasizes a continuous concern of low-level hardware (\gps), and \S\ref{rq3} highlights that the majority of questions posed by practitioners are related to the energy consumption of hardware accessories, such as Bluetooth and GPS.
Specifically, a software application that frequently makes use of Bluetooth or GPS can be tuned to activate the hardware accessories only when necessary rather than keeping them on continuously. 
\Revision{For example, if coarse accuracy of position is acceptable, practitioners could call \code{requestLocationUpdates()} in Android's \code{LocationManager} with a large minimum time interval and \code{NETWORK$\_$PROVIDER} (which relies on cell-tower/Wi-Fi rather than GPS) to avoid unnecessary high-energy GPS fixes (\sopost{3034890}~\cite{so3034890}).}



\Revision{\textbf{Leveraging platform's built-in scheduling and power management APIs to minimize unnecessary background tasks.}}
As discussed in \S\ref{rq3}, mobile OS like Android have significant energy concerns, \Revision{whereas {\tt Java} is the most concerned single \pl for Android.
We observe that excessive background tasks, such as tight wakelock loops, frequent sensor listeners, and unmanaged threads, keep CPU and other hardware active, which rapidly drain the battery.
To mitigate this, practitioners could use \code{AlarmManager} instead of a manual wakelock loop to save power at idle time.
For example, tasks can be scheduled with an exact alarm 
\code{setExactAndAllowWhileIdle()} that fires even under the idle mode, allowing the system to sleep fully until that moment instead of constantly holding CPU awake (e.g., \sopost{10468305}~\cite{so10468305}).
Besides, practitioners could hold the \code{WakeLock} in persistent fields rather than locals so they can reliably release it and let CPU return to idle sooner. 
For example, \code{WakeLock} could be used as a class-level (non-local) field to ensure that the lock is not garbage-collected immediately (e.g., \sopost{38325958}~\cite{so38325958}).}

\Revision{\textbf{Making optimized usage of 
built-in API routines, especially being precautionary of per-element processing patterns.
}
In \S\ref{rq2}, our findings suggest that {\tt API USAGE} issues remain consistently significant ($\sim$30\%) across all years.
We observe that such issues typically stem from per‐element processing patterns (e.g. nested loops or full‐list rebuilds) and by mismanaging asynchronous or lifecycle APIs, which lead to repeated work or busy‐wait loops that spike CPU usage and power draw.
For instance, in Kivy GUI updates (\sopost{66972381}~\cite{so66972381}), rather than reconstructing the full RecycleView dataset on each incoming serial-port message,
practitioners are suggested to keep a persistent Python list of messages, append only the new entries to it, and then update the RecycleView’s \code{data} property only when necessary to avoid a full list rebuild and drastically cutting CPU and power usage.
}

\Revision{
\textbf{Pushing data‐handling optimizations down to the level where they can actually control resource usage (e.g., OS/kernel, hardware components, and data structures) rather than trying to tweak high-level app code.}
As discussed in \S\ref{rq1}, we notice a technological shift to an emphasis on data-centric topics (e.g., \data).
}
In fact, questions about \data not only have the highest non-acceptance rates but also a longer median response time.
\Revision{To address this challenge, practitioners could dynamically cache power-gating via dynamic voltage and frequency scaling---by lowering the supply voltage and clock rate of cache subarrays and ramping up or powering on the data banks only when a tag lookup hits---to conserve power on workloads with sparse cache traffic.
For example, \sopost{41749881}~\cite{so41749881} suggests that physically separating the tag and data arrays to keep the tag array powered at a low voltage/frequency and only power up the data banks on a cache‐hit, thereby saving energy on workloads with low cache‐access intensity.}

%% file: Text/Threats.tex
\section{Threats to Validity}\label{sec:TV}


\MyPara{Construct Validity}
Even though our data cover a wide range of SO posts, our selection of posts may not be exhaustive. 
The automatic keyword filtering may missed some relevant posts, but manually verifying all posts in the data dump is infeasible.
The size of the data and the linguistic freedom of presentation---different words present the same meaning---can have a significant impact on the results.
To mitigate these threats, we conduct manual inspection twice: one is to filter out the false positive posts (matching keywords but unrelated to energy consumption), and another is to validate the LDA result and intention categories. The manual inspection is performed by specialists (two PhD researchers in Software Engineering with over 7 years of experience) following Seaman's closed coding procedures~\citep{Seaman} (\S\ref{identifyPost}).
\Revision{While analyzing questions and text content may provide deeper insights (\S\ref{rq3}), we rely on SO tags as they
are community-curated 
to reflect the primary technologies or topics involved. Leveraging tags allows our analysis to remain consistent and replicable across large datasets. Meanwhile, prior studies~\cite{Pinto, Barua} demonstrate that SO tags accurately reflect developers' primary technological concerns. To mitigate limitations from tag-based analysis, we merge synonyms or  related tags, thereby capturing broader trends and technology categories comprehensively to ensure our results represent the underlying data.}

\MyPara{Internal Validity}
We choose the LDA model as our research method, whose results may be inaccurate. To mitigate this, the LDA topics are further validated via manual inspection. 
Another challenge arises from employing LLMs to label intents in RQ2. 
To ensure that the LLMs robustly generate accurate and contextually relevant outputs, we follow best practices from previous studies~\cite{fathollahzadeh25towards, triem24tipping, khan24debating} and engineered prompts~\cite{openai-prompt, huggingface-prompt} and also manually verify the results are valid.
We select the tags of posts to summarize technology categories with regard to the occurrences of the tags. 
Nonetheless, the occurrence may not indicate the true relationship between the theme of a post and the technologies.
To mitigate this, we cover the most development category-related tags, which increases our relevance to the study.
\Revision{We utilize LLM voting to infer question intentions, which might be unreliable if not carefully configured. 
To mitigate this, we manually label intents and compare them with LLM-voted results. 
The result shows a substantial consensus (Cohen’s Kappa of 0.64) between human work and LLMs.}

\noindent\textbf{External Validity.}
Our results apply only to questions interested in energy usage on SO. It does not include other Q\&A websites, such as Ask Ubuntu, nor does it include the posts asked in other languages (e.g., French). Although SO is the most popular development Q\&A website, further investigation could be conducted to subsume additional sources.

%% file: Text/Concl.tex
\section{Conclusion}\label{sec:Conclusion}

In this paper, we conduct an empirical study to investigate how practitioners on Stack Overflow are concerned about energy consumption.
By studying 1,193 energy-related questions, we observe the following:
(1)~understanding the energy impact of \gps service and \resource used by their programs are of particular important to practitioners; 
(2)~\polling and \data are the most difficult topics to receive community support due to in-depth knowledge required;
(3)~most practitioners struggle with fundamental energy concepts due to complex API documentation and practical experience required;
and (4)~practitioners consider energy-efficient development from multiple levels---\os, \hw, and \pl, where {\tt Android}, {\tt Accessory} , and {\tt Java} are the most concerned technologies, respectively.
Based on our findings, we provide actionable suggestions for knowledge-sharing communities and practitioners.

%% file: bib.bib
@article{category,
author = {Beyer, Stefanie and Macho, Christian and Di Penta, Massimiliano and Pinzger, Martin},
title = {What Kind of Questions Do Developers Ask on Stack Overflow? A Comparison of Automated Approaches to Classify Posts into Question Categories},
year = {2020},
issue_date = {May 2020},
publisher = {Kluwer Academic Publishers},
address = {USA},
volume = {25},
number = {3},
issn = {1382-3256},
journal = {Empirical Softw. Engg.},
month = {May},
pages = {2258–2301},
numpages = {44}
}

@article{Yang,
author = {Yang, Xin-Li and Lo, David and Xia, Xin and Wan, Zhiyuan and Sun, Jian-Ling},
year = {2016},
month = {September},
pages = {910-924},
title = {What Security Questions Do Developers Ask? A Large-Scale Study of Stack Overflow Posts},
volume = {31},
journal = {Journal of Computer Science and Technology}
}

@inproceedings{Bagherzadeh,
author = {Bagherzadeh, Mehdi and Khatchadourian, Raffi},
title = {Going Big: A Large-Scale Study on What Big Data Developers Ask},
year = {2019},
isbn = {9781450355728},
publisher = {Association for Computing Machinery},
address = {New York, NY, USA},
booktitle = {Proceedings of the 2019 27th ACM Joint Meeting on European Software Engineering Conference and Symposium on the Foundations of Software Engineering},
pages = {432–442},
numpages = {11},
location = {Tallinn, Estonia},
series = {ESEC/FSE 2019}
}

@article{Barua,
author = {Barua, Anton and Thomas, Stephen W. and Hassan, Ahmed E.},
title = {What Are Developers Talking about? An Analysis of Topics and Trends in Stack Overflow},
year = {2014},
publisher = {Kluwer Academic Publishers},
address = {USA},
volume = {19},
number = {3},
issn = {1382-3256},
journal = {Empirical Software Engineering},
month = {June},
pages = {619–654},
numpages = {36},
}

@ARTICLE{blockchain,  author={Wan, Zhiyuan and Xia, Xin and Hassan, Ahmed E.},  journal={IEEE Transactions on Software Engineering},   title={What Do Programmers Discuss About Blockchain? A Case Study on the Use of Balanced LDA and the Reference Architecture of a Domain to Capture Online Discussions About Blockchain Platforms Across Stack Exchange Communities},   year={2021},  volume={47},  number={7},  pages={1331-1349}}

@article{Griffiths,
	author = {Thomas L. Griffiths and Mark Steyvers and Joshua B. Tenenbaum},
	title = {Topics in Semantic Representation},
	volume = {114},
	pages = {211--244},
	number = {2},
	journal = {Psychological Review},
	year = {2007}
}

@inproceedings{rehurek,
      title = {{Software Framework for Topic Modelling with Large Corpora}},
      author = {Radim {\v R}eh{\r u}{\v r}ek and Petr Sojka},
      booktitle = {{Proceedings of the LREC 2010 Workshop on New
           Challenges for NLP Frameworks}},
      pages = {45--50},
      year = 2010,
      month = May,
      day = 22,
      publisher = {ELRA},
      address = {Valletta, Malta},
      language={English}
}

@inproceedings{Greene,
author = {Greene, Derek and O'Callaghan, Derek and Cunningham, Padraig},
year = {2014},
month = {April},
pages = {498-513},
title = {How Many Topics? Stability Analysis for Topic Models},
isbn = {978-3-662-44847-2},
}

@article{Agrawal,
title = {What is wrong with topic modeling? And how to fix it using search-based software engineering},
journal = {Information and Software Technology},
volume = {98},
pages = {74-88},
year = {2018},
issn = {0950-5849},
author = {Amritanshu Agrawal and Wei Fu and Tim Menzies},
}

@ARTICLE{Seaman,
  author={Seaman, C.B.},
  journal={IEEE Transactions on Software Engineering}, 
  title={Qualitative methods in empirical studies of software engineering}, 
  year={1999},
  volume={25},
  number={4},
  pages={557-572}}

@inproceedings{Newman,
  title={Automatic evaluation of topic coherence},
  author={ Newman, David  and  Lau, Jey Han  and  Grieser, Karl  and  Baldwin, Timothy },
  booktitle={Human Language Technologies: Conference of the North American Chapter of the Association of Computational Linguistics},
  year={2010},
}

@article{Roder,
  title={Exploring the Space of Topic Coherence Measures},
  author={ Röder, Michael  and  Both, Andreas  and  Hinneburg, Alexander },
  journal={ACM},
  pages={399-408},
  year={2015},
}

@inproceedings{Pinto,
author = {Pinto, Gustavo and Castor, Fernando and Liu, Yu David},
title = {Mining Questions about Software Energy Consumption},
year = {2014},
isbn = {9781450328630},
publisher = {Association for Computing Machinery},
address = {New York, NY, USA},
booktitle = {Proceedings of the 11th Working Conference on Mining Software Repositories},
pages = {22–31},
numpages = {10},
location = {Hyderabad, India},
series = {MSR 2014}
}

@ARTICLE {Pang,
author = {C. Pang and A. Hindle and B. Adams and A. E. Hassan},
journal = {IEEE Software},
title = {What Do Programmers Know about Software Energy Consumption?},
year = {2016},
volume = {33},
number = {03},
issn = {1937-4194},
pages = {83-89},
publisher = {IEEE Computer Society},
address = {Los Alamitos, CA, USA},
month = {May}
}

@inproceedings{Manotas,
author = {Manotas, Irene and Bird, Christian and Zhang, Rui and Shepherd, David and Jaspan, Ciera and Sadowski, Caitlin and Pollock, Lori and Clause, James},
title = {An Empirical Study of Practitioners' Perspectives on Green Software Engineering},
year = {2016},
isbn = {9781450339001},
publisher = {Association for Computing Machinery},
address = {New York, NY, USA},
booktitle = {Proceedings of the 38th International Conference on Software Engineering},
pages = {237–248},
numpages = {12},
location = {Austin, Texas},
series = {ICSE '16}
}

@INPROCEEDINGS{Venkatesh,  author={Venkatesh, Pradeep K. and Wang, Shaohua and Zhang, Feng and Zou, Ying and Hassan, Ahmed E.},  booktitle={2016 IEEE International Conference on Web Services (ICWS)},   title={What Do Client Developers Concern When Using Web APIs? An Empirical Study on Developer Forums and Stack Overflow},   year={2016},  volume={},  number={},  pages={131-138}}

@article{Blei,
author = {Blei, David M. and Ng, Andrew Y. and Jordan, Michael I.},
title = {Latent Dirichlet Allocation},
year = {2003},
issue_date = {3/1/2003},
publisher = {JMLR.org},
volume = {3},
number = {null},
issn = {1532-4435},
journal = {J. Mach. Learn. Res.},
month = Mar,
pages = {993–1022},
numpages = {30}
}

@article{2030,
author = {Andrae, Anders},
year = {2020},
month = {June},
pages = {19-31},
title = {New perspectives on internet electricity use in 2030},
volume = {3},
journal = {Engineering and Applied Science Letters}
}

@article{emissions,
author = {Freitag, Charlotte and Berners-Lee, Mike and Widdicks, Kelly and Knowles, Bran and Blair, Gordon and Friday, Adrian},
year = {2021},
month = {September},
pages = {100340},
title = {The real climate and transformative impact of ICT: A critique of estimates, trends, and regulations},
volume = {2},
journal = {Patterns}
}

@article{software1,
author = {Ciancarini, Paolo and Ergasheva, Shokhista and Zamira, Kholmatova and Kruglov, Artem and Succi, Giancarlo and Vasquez, Xavier and Zuev, Evgeniy},
year = {2020},
month = {October},
pages = {1678},
title = {Analysis of Energy Consumption of Software Development Process Entities},
volume = {9},
journal = {Electronics}
}

@article{Jin,
author = {Jin, Lei and Duan, Keran and Tang, Xu},
year = {2018},
month = {01},
pages = {145},
title = {What Is the Relationship between Technological Innovation and Energy Consumption? Empirical Analysis Based on Provincial Panel Data from China},
volume = {10},
journal = {Sustainability}
}

@article{Trends2030,
author = {Wang, Peng and zhong, peirong and Yu, Min and Pu, Yanru and Zhang, Shuainan and Yu, Ping},
year = {2022},
month = {July},
pages = {},
title = {Trends in energy consumption under the multi-stage development of ICT: Evidence in China from 2001 to 2030},
volume = {8},
journal = {Energy Reports}
}

@ARTICLE{software2,
  author={Verdecchia, Roberto and Lago, Patricia and Ebert, Christof and de Vries, Carol},
  journal={IEEE Software}, 
  title={Green IT and Green Software}, 
  year={2021},
  volume={38},
  number={6},
  pages={7-15}}

@misc{data2030, 
title={How green is your software?}, 
journal={Harvard Business Review}, 
author={Podder, Sanjay and Burden, Adam and Singh, Shalabh Kumar and Maruca, Regina}, 
year={2020}, 
month={September},}

@inproceedings{Allamanis, 
  author = {Allamanis, Miltiadis and Sutton, Charles}, 
  title = {Why, When, and What: Analyzing Stack Overflow Questions by Topic, Type, and Code}, 
  year = {2013}, 
  booktitle = {Proceedings of the 10th Working Conference on Mining Software Repositories}, 
  pages = {53–56}, 
}

@inproceedings{LDA,
author = {Blei, David and Ng, Andrew and Jordan, Michael},
year = {2001},
month = {Jan.},
pages = {601-608},
title = {Latent Dirichlet Allocation},
volume = {3},
journal = {The Journal of Machine Learning Research}
}

@article{RS,
  title={What are mobile developers asking about? a large scale study using stack overflow},
  author={Rosen, Christoffer and Shihab, Emad},
  journal={Empirical Software Engineering},
  volume={21},
  number={3},
  pages={1192--1223},
  year={2016},
  publisher={Springer}
}

@article{kappa,
author = {Jacob Cohen},
title ={A Coefficient of Agreement for Nominal Scales},
journal = {Educational and Psychological Measurement},
volume = {20},
number = {1},
pages = {37-46},
year = {1960}
}

@misc{so19722950,
    Title = {Do sse instructions consume more power/energy?},
    Author = {Antonio},
    url = {https://stackoverflow.com/questions/19722950},
    year = {2013}
}

@misc{so13476223,
    Title = {Optimising an 1D heat equation using SIMD},
    Author = {Lags Moomba},
    url ={https://stackoverflow.com/questions/13476223},
    year = {2012}
}

@misc{so61882,
    Title = {Power Efficient Software Coding},
    Author = {golden},
    url ={https://stackoverflow.com/questions/61882},
    year = {2008}
}

@misc{so66697291,
    Title = {Measure energy consumed by a python script in Raspberry Pi (Raspbian)},
    Author = {cesarcastellon.sec},
    url ={https://stackoverflow.com/questions/66697291},
    year = {2021}
}

@misc{so42850419,
    Title = {My app keep the device awake all the time},
    Author = {Amit Shadadi},
    url ={https://stackoverflow.com/questions/42850419},
    year = {2017}
}

@misc{so38158828,
    Title = {Android Google Play Service: Best practices to make location request},
    Author = {user3290180},
    url ={https://stackoverflow.com/questions/38158828},
    year = {2016}
}

@misc{so41749881,
    Title = {Separated tag array versus combined with data array},
    Author = {Mrchacha},
    url = {https://stackoverflow.com/questions/41749881},
    year = {2017}
}

@misc{so9863131,
    Title = {Android accelerometer, sensor usage and power consumption},
    Author = {gorn},
    url = {https://stackoverflow.com/questions/9863131},
    year = {2012}
}

@misc{so3034890,
    Title = {Determine Android phone's proximity to known point while conserving power},
    Author = {ahsteele},
    url = {https://stackoverflow.com/questions/3034890},
    year = {2010}
}

@misc{so10920904,
    Title = {Energy efficient GPS tracking},
    Author = {Prashant Tiwari},
    url = {https://stackoverflow.com/questions/10920904},
    year = {2012}
}

@misc{so58156084,
    Title = {Is there a way to limit the amount of CPU a c++ application uses},
    Author = {programmerRaj},
    url = {https://stackoverflow.com/questions/58156084},
    year = {2019}
}

@misc{so5765212,
    Title ={Is it ok to update a widget frequently for a short period of time?},
    Author = {olivierg},
    url = {https://stackoverflow.com/questions/5765212},
    year = {2011}
}

@misc{so61884014,
    Title = {How do I make sure that a Java thread will never run an infinite loop?},
    Author = {Alejandro Servetto},
    url = {https://stackoverflow.com/questions/61884014},
    year = {2020}
}

@misc{so1738515,
Title = {RAM memory reallocation - Windows and Linux},
    Author = {kjv},
    url = {https://stackoverflow.com/questions/1738515},
    year = {2009}
}

@misc{so53173447,
  title = {Optimize power consumption with STM32L4 ADC},
  author = {Guillaume Petitjean},
  url = {https://stackoverflow.com/questions/53173447},
    year = {2018}
}

@misc{so17134522,
Title = {How to compare the performance of Android Apps written in Java and Xamarin C\#? Anyway to check quantitative data (code \& results)},
    Author = {gregko},
    url = {https://stackoverflow.com/questions/17134522},
    year = {2013}
}

@misc{so5151872,
Title = {How can I sense if a phone is in standby or sleep mode(dor Nokia)?},
    Author = {atasoyh},
    url = {https://stackoverflow.com/questions/5151872},
    year = {2011}
}

@misc{so53359800,
Title = {Java TCP server for communicating with an IoT device},
    Author = {Iman H},
url= {https://stackoverflow.com/questions/53359800},
    year = {2018}
}

@misc{so724349,
Title = {How can I measure the energy consumption of my application on Windows Mobile and Windows CE?},
    Author = {Daniel Rikowski},
    url = {https://stackoverflow.com/questions/724349},
    year = {2009}
}

@misc{so2902382,
Title = {Android Nexus One - Can I save energy with color scheme?},
    Author = {Maksym Gontar},
    url = {https://stackoverflow.com/questions/2902382},
    year = {2010}
}

@misc{so38635068,
Title = {Does Android have a different behavior for WakeLock when the device is connected to power source?},
    Author = {Luis Cruz},
    url = {https://stackoverflow.com/questions/38635068},
    year = {2016}
}

@misc{so51973350,
Title = {windowed OpenGL first frame delay after idle},
    Author = {quantumfoam},
    url = {https://stackoverflow.com/questions/51973350},
    year = {2018}
}

@misc{so15698999,
Title = {Implementing a battery widget with alarmservice?},
    Author = {awonder},
    url = {https://stackoverflow.com/questions/15698999},
    year = {2013}
}

@misc{so23617388,
Title = {How to select element in a php multi dimentional array like the WHERE in mysql},
    Author = {bob dylan},
    url = {https://stackoverflow.com/questions/23617388},
    year = {2014}
}

@misc{so54028927,
Title = {Implementing a battery widget with alarmservice?},
    Author = {Jovan Pacheco},
    url = {https://stackoverflow.com/questions/54028927},
    year = {2019}
}

@misc{so53577434,
Title = {OpenGL: How to minimize drawing?},
    Author = {popoe},
    url = {https://stackoverflow.com/questions/53577434},
    year = {2018}
}

@misc{so28003660,
Title = {Why does reading the CP15 c1 control register fail (ARMv7 inline asm)?},
    Author = {wondering},
    url = {https://stackoverflow.com/questions/28003660},
    year = {2015}
}

@misc{so44228005,
Title = {How to measure power consumption of Jetson TX1?},
    Author = {MinsubKim},
    url = {https://stackoverflow.com/questions/44228005},
    year = {2017}
}

@misc{so4485153,
Title = {Estimating process energy usage on PCs (x86)},
    Author = {Eitan},
    url = {https://stackoverflow.com/questions/4485153},
    year = {2010}
}

@misc{so29616977,
Title = {Getting latitude and longitude updates based off of change in distance for Android},
    Author = {Jamaal},
    url = {https://stackoverflow.com/questions/29616977},
    year = {2015}
}

@misc{so43596157,
Title = {Detect car parking efficiently on Android},
    Author = {Eitan},
    url = {https://stackoverflow.com/questions/43596157},
    year = {2017}
}

@misc{so23560949,
Title = {Android LocationListener for a certain time},
    Author = {dev\_android},
    url = {https://stackoverflow.com/questions/23560949},
    year = {2014}
}

@misc{so26336225,
Title = {Save continuous location data on server through android},
    Author = {Manish Bane},
    url = {https://stackoverflow.com/questions/26336225},
    year = {2014}
}

@misc{so8782922,
Title = {Decentralized Clustering library for Java},
    Author = {Codifier},
    url = {https://stackoverflow.com/questions/8782922},
    year = {2012}
}

@misc{so50064364,
Title = {Power consumption of a Systick timer on STM32},
    Author = {Guillaume Petitjean},
    url = {https://stackoverflow.com/questions/50064364},
    year = {2018}
}

@misc{so28760898,
Title = {Power consumption of a Systick timer on STM32},
    Author = {user534498},
    url = {https://stackoverflow.com/questions/28760898},
    year = {2015}
}

@misc{so27383269,
Title = {Why dynamic power consumption is always zero?},
    Author = {Guillaume Petitjean},
    url = {https://stackoverflow.com/questions/27383269},
    year = {2014}
}

@misc{so39804144,
Title = {network-on-chip verilog code},
    Author = {maryam},
    url = {https://stackoverflow.com/questions/39804144},
    year = {2016}
}

@misc{so11366329,
Title = {Implementing an event and periodically driven "script language" in C++?},
    Author = {Chris},
    url = {https://stackoverflow.com/questions/11366329},
    year = {2012}
}

@misc{so1298600,
Title = {Bluetooth UUID discovery},
    Author = {Antony Carthy},
    url = {https://stackoverflow.com/questions/1298600},
    year = {2009}
}

@misc{so3655806,
Title = {How to measure power consumed by my algorithm?},
    Author = {Ciro Santilli OurBigBook.com},
    url = {https://stackoverflow.com/questions/3655806},
    year = {2010}
}

@misc{so6051807,
Title = {per process power consumption in Android},
    Author = {skaffman},
    url = {https://stackoverflow.com/questions/6051807},
    year = {2011}
}

@misc{so6553595,
Title = {Sending a String array through a tigase server from one Android to another using XMPP protocol},
    Author = {Nimantha},
    url = {https://stackoverflow.com/questions/6553595},
    year = {2011}
}

@misc{so13873570,
Title = {Correct way to implement a stock price alert system},
    Author = {Squonk},
    url = {https://stackoverflow.com/questions/13873570},
    year = {2012}
}

@misc{so31368757,
Title = {How to obtain the remaining energy from BatteryManager.BATTERY\_PROPERTY\_ENERGY\_COUNTER},
Author = {Spirrow},
url = {https://stackoverflow.com/questions/31368757},
    year = {2015}
}

@misc{so8100506,
Title = {How do I use Android PowerProfile private API?},
    Author = {halfer},
    url = {https://stackoverflow.com/questions/8100506},
    year = {2011}
}

@misc{so11398732,
Title = {How do I receive the system broadcast when GPS status has changed?},
    Author = {dda},
    url = {https://stackoverflow.com/questions/11398732},
    year = {2012}
}

@misc{so49202065,
Title = {How to get CPU power consumption in PowerShell?},
    Author = {lel},
    url = {https://stackoverflow.com/questions/49202065},
    year = {2018}
}

@misc{so7312597,
Title = {Is using hardware performance counters a good idea},
    Author = {MetallicPriest},
    url = {https://stackoverflow.com/questions/7312597},
    year = {2011}
}

@misc{so410122,
Title = {Reducing power consumption on a low-load server running WinXP?},
    Author = {Ksempac},
    url = {https://stackoverflow.com/questions/410122},
    year = {2009}
}

@misc{so72711521,
Title = {Win11 22H2 Programmatically switch between predefined power modes (Best power efficiency/Balanced/Best performance)},
    Author = {Jun Ish},
    url = {https://stackoverflow.com/questions/72711521},
    year = {2022}
}

@misc{so40090627,
Title = {How to structure a notification system for a chat app using Firebase Database and Firebase Notification},
    Author = {Jackie Degl'Innocenti },
    url = {https://stackoverflow.com/questions/40090627},
    year = {2016}
}

@misc{so48766582,
Title = {Is there a way to stop Qt rendering temporarily?},
    Author = {
Étienne},
    url = {https://stackoverflow.com/questions/48766582},
    year = {2018}
}

@misc{so66972381,
Title = {Kivy Desktop App for Serial Data communication in fastest possible way},
    Author = {
SharonWingle},
    url = {https://stackoverflow.com/questions/66972381},
    year = {2021}
}

@misc{so10468305,
Title = {Reducing power consumption},
    Author = {
Zelig},
    url = {https://stackoverflow.com/questions/10468305},
    year = {2012}
}

@misc{so38325958,
Title = {android service logging sensor data uninterrupted},
    Author = {
apidae},
    url = {https://stackoverflow.com/questions/38325958},
    year = {2016}
}

@misc{speedstep,
Title = {Overview of Enhanced Intel SpeedStep® Technology for Intel® Processors},
    Author = {Intel},
    url = {https://www.intel.com/content/www/us/en/support/articles/000007073/processors.html},
    year = {2022}
}

@misc{quiet,
  author = {AMD},
  title = {Cool `n' Quiet™ Technology Installation Guide for AMD Athlon™ 64 Processor Based Systems},
  url={https://web.archive.org/web/20070409045621/http://www.amd.com/us-en/assets/content_type/DownloadableAssets/Cool_N_Quiet_Installation_Guide3.pdf},
  year = {2004},
  month = {June}
}

@misc{macKernal,
  author = {Apple},
  title = {Kernel Programming Guide},
  url = {https://developer.apple.com/library/archive/documentation/Darwin/Conceptual/KernelProgramming/performance/performance.html#//apple_ref/doc/uid/TP30000905-CH207-BEHJDFCA},
  year = {2013},
  month = {08}
}

@misc{scheduling,
Title = {Scheduling},
    Author = {MicroSoft},
    url = {https://learn.microsoft.com/en-us/windows/win32/procthread/scheduling},
year={2021}
}

@misc{4o,
      title={GPT-4 Technical Report}, 
      author={OpenAI},
      year={2024},
      eprint={2303.08774},
      archivePrefix={arXiv},
      primaryClass={cs.CL},
      url={https://arxiv.org/abs/2303.08774}, 
}

@misc{qwq32b,
    title = {QwQ-32B: Embracing the Power of Reinforcement Learning},
    url = {https://qwenlm.github.io/blog/qwq-32b/},
    author = {Qwen Team},
    month = {March},
    year = {2025}
}

@misc{openai-prompt,
    title = {Prompt engineering},
    url = {https://platform.openai.com/docs/guides/prompt-engineering},
    author = {OpenAI}
}

@misc{huggingface-prompt,
    title = {Chat Templates},
    url = {https://huggingface.co/docs/transformers/main/en/chat_templating},
    author = {Huggingface}
}

@misc{deepseekr1,
      title={DeepSeek-R1: Incentivizing Reasoning Capability in LLMs via Reinforcement Learning}, 
      author={DeepSeek-AI},
      year={2025},
      eprint={2501.12948},
      archivePrefix={arXiv},
      primaryClass={cs.CL}
}

@inproceedings{Wallach09LDAparam,
  title={Rethinking LDA: Why Priors Matter},
  author={Hanna M. Wallach and David Mimno and Andrew McCallum},
  booktitle={Neural Information Processing Systems},
  year={2009}
}

@inproceedings{Zeng02OS,
author = {Zeng, Heng and Ellis, Carla S. and Lebeck, Alvin R. and Vahdat, Amin},
title = {ECOSystem: managing energy as a first class operating system resource},
year = {2002},
isbn = {1581135742},
publisher = {Association for Computing Machinery},
address = {New York, NY, USA},
booktitle = {Proceedings of the 10th International Conference on Architectural Support for Programming Languages and Operating Systems},
pages = {123–132},
numpages = {10},
location = {San Jose, California},
series = {ASPLOS X}
}

@inproceedings{Carroll10phone,
author = {Carroll, Aaron and Heiser, Gernot},
title = {An analysis of power consumption in a smartphone},
year = {2010},
publisher = {USENIX Association},
address = {USA},
booktitle = {Proceedings of the 2010 USENIX Conference on USENIX Annual Technical Conference},
pages = {21},
numpages = {1},
location = {Boston, MA},
series = {USENIXATC'10}
}

@INPROCEEDINGS{Chen11Map,
  author={Chen, Yanpei and Ganapathi, Archana and Griffith, Rean and Katz, Randy},
  booktitle={2011 IEEE 19th Annual International Symposium on Modelling, Analysis, and Simulation of Computer and Telecommunication Systems}, 
  title={The Case for Evaluating MapReduce Performance Using Workload Suites}, 
  year={2011},
  volume={},
  number={},
  pages={390-399}}

@book{benini2011compilers,
  title={Compilers and Operating Systems for Low Power},
  author={Benini, Luca and Kandemir, Mahmut and Ramanujam, J},
  year={2011},
  publisher={Springer Science \& Business Media}
}

@misc{LDApass,
    Title = {LDA Model},
    Author = {
Gensim},
    url = {https://radimrehurek.com/gensim/auto_examples/tutorials/run_lda.html}
}

@book{marinescu2022cloud,
  title={Cloud computing: theory and practice},
  author={Marinescu, Dan C},
  year={2022},
  publisher={Morgan Kaufmann}
}

@incollection{Zelkowitz11Cloud,
title = {Chapter 3 - A Taxonomy and Survey of Energy-Efficient Data Centers and Cloud Computing Systems},
editor = {Marvin V. Zelkowitz},
series = {Advances in Computers},
publisher = {Elsevier},
volume = {82},
pages = {47-111},
year = {2011},
issn = {0065-2458},
author = {Anton Beloglazov and Rajkumar Buyya and Young Choon Lee and Albert Zomaya},
}

@INPROCEEDINGS{Bangash21Location,
  author={Bangash, Abdul Ali and Tiganov, Daniil and Ali, Karim and Hindle, Abram},
  booktitle={2021 IEEE International Conference on Software Maintenance and Evolution (ICSME)}, 
  title={Energy Efficient Guidelines for iOS Core Location Framework}, 
  year={2021},
  volume={},
  number={},
  pages={320-331}
}

@INPROCEEDINGS{Moura15MiningEnergy,
  author={Moura, Irineu and Pinto, Gustavo and Ebert, Felipe and Castor, Fernando},
  booktitle={2015 IEEE/ACM 12th Working Conference on Mining Software Repositories}, 
  title={Mining Energy-Aware Commits}, 
  year={2015},
  volume={},
  number={},
  pages={56-67}}

@INPROCEEDINGS{Bao16Android,
  author={Bao, Lingfeng and Lo, David and Xia, Xin and Wang, Xinyu and Tian, Cong},
  booktitle={2016 IEEE/ACM 13th Working Conference on Mining Software Repositories (MSR)}, 
  title={How Android App Developers Manage Power Consumption? - An Empirical Study by Mining Power Management Commits}, 
  year={2016},
  volume={},
  number={},
  pages={37-48},
  doi={}}

@article{Cruz19Catelog,
author = {Cruz, Luis and Abreu, Rui},
title = {Catalog of energy patterns for mobile applications},
year = {2019},
issue_date = {August    2019},
publisher = {Kluwer Academic Publishers},
address = {USA},
volume = {24},
number = {4},
issn = {1382-3256},
journal = {Empirical Softw. Engg.},
month = aug,
pages = {2209–2235},
numpages = {27}
}

@Inbook{Dragoni17Microservices,
author="Dragoni, Nicola
and Giallorenzo, Saverio
and Lafuente, Alberto Lluch
and Mazzara, Manuel
and Montesi, Fabrizio
and Mustafin, Ruslan
and Safina, Larisa",
LONGeditor="Mazzara, Manuel
and Meyer, Bertrand",
title="Microservices: Yesterday, Today, and Tomorrow",
bookTitle="Present and Ulterior Software Engineering",
year="2017",
publisher="Springer International Publishing",
address="Cham",
pages="195--216",
isbn="978-3-319-67425-4"
}

@article{gomez12BLEoverview,
  title={Overview and evaluation of bluetooth low energy: An emerging low-power wireless technology},
  author={Gomez, Carles and Oller, Joaquim and Paradells, Josep},
  journal={sensors},
  volume={12},
  number={9},
  pages={11734--11753},
  year={2012},
  publisher={Molecular Diversity Preservation International (MDPI)}
}

@article{Aouedi24IoT,
  author={Aouedi, Ons and Vu, Thai-Hoc and Sacco, Alessio and Nguyen, Dinh C. and Piamrat, Kandaraj and Marchetto, Guido and Pham, Quoc-Viet},
  journal={IEEE Communications Surveys \& Tutorials}, 
  title={A Survey on Intelligent Internet of Things: Applications, Security, Privacy, and Future Directions}, 
  year={2024},
  volume={},
  number={},
  pages={1-1}}

@article{triem24tipping,
  title={“Tipping the Balance”: Human Intervention in Large Language Model Multi-Agent Debate},
  author={Triem, Haley and Ding, Ying},
  journal={Proceedings of the Association for Information Science and Technology},
  volume={61},
  number={1},
  pages={361--373},
  year={2024},
  publisher={Wiley Online Library}
}

@article{khan24debating,
  title={Debating with more persuasive llms leads to more truthful answers},
  author={Khan, Akbir and Hughes, John and Valentine, Dan and Ruis, Laura and Sachan, Kshitij and Radhakrishnan, Ansh and Grefenstette, Edward and Bowman, Samuel R and Rockt{\"a}schel, Tim and Perez, Ethan},
  journal={arXiv preprint arXiv:2402.06782},
  year={2024}
}

@article{fathollahzadeh25towards,
  title={Towards Refining Developer Questions using LLM-Based Named Entity Recognition for Developer Chatroom Conversations},
  author={Fathollahzadeh, Pouya and Mezouar, Mariam El and Li, Hao and Zou, Ying and Hassan, Ahmed E},
  journal={arXiv preprint arXiv:2503.00673},
  year={2025}
}

@INPROCEEDINGS{JUGRAN21Spacy,
  author={Jugran, Swaranjali and Kumar, Ashish and Tyagi, Bhupendra Singh and Anand, Vivek},
  booktitle={2021 International Conference on Advance Computing and Innovative Technologies in Engineering (ICACITE)}, 
  title={Extractive Automatic Text Summarization using SpaCy in Python \& NLP}, 
  year={2021},
  volume={},
  number={},
  pages={582-585}}

@article{Hu2022DataDI,
  title={Data driven identification of international cutting edge science and technologies using SpaCy},
  author={Chunqi Hu and Huaping Gong and Yiqing He},
  journal={PLoS ONE},
  year={2022},
  volume={17}
}

@inproceedings{Zhuang10gps,
author = {Zhuang, Zhenyun and Kim, Kyu-Han and Singh, Jatinder Pal},
title = {Improving energy efficiency of location sensing on smartphones},
year = {2010},
isbn = {9781605589855},
publisher = {Association for Computing Machinery},
address = {New York, NY, USA},
booktitle = {Proceedings of the 8th International Conference on Mobile Systems, Applications, and Services},
pages = {315–330},
numpages = {16},
location = {San Francisco, California, USA},
series = {MobiSys '10}
}

@inproceedings{Siemers25OS,
author = {Wander Siemers and Luís Cruz and Mattia Fazzini},
title = {Toward Understanding and Detecting Battery Saver Issues in Android Apps},
year = {2025},
publisher = {Association for Computing Machinery},
address = {New York, NY, USA},
booktitle = {Proceedings of the 12th International Conference on Mobile Software Engineering and Systems},
location = {Ottawa, Canada},
series = {MOBILESoft '25}
}

@online{RepPackage,
  author = {{Bihui Jin}},
  title = {Replication Package},
  year = 2025,
  url = {https://github.com/Bihui-Jin/Suppmaterial-ICSE26-Energy-Efficient-Software-Development}
}
